%% file: main.tex
\documentclass[12pt]{article}
\usepackage[utf8x]{inputenc}
\usepackage{geometry}
\usepackage[T1]{fontenc}
\usepackage{graphicx}
\usepackage{caption}
\usepackage{subcaption}
\usepackage{comment}
\usepackage{scalerel}
\DeclareMathSizes{12}{12}{8}{2}
\usepackage[symbol]{footmisc}

\usepackage[most]{tcolorbox}
\usepackage{xcolor}
\newcommand{\CD}[1]{\color{blue}{ CD: \bf #1}\color{black}}
\newcommand{\OWWW}{\mathcal{O}_{\widetilde{W}WW}}
\newcommand{\obothi}{\hspace{-1mm}\begin{array}{c}\quad\\[-8mm]\scriptscriptstyle{\left({\wedge}\right)\,}\\[-1.7mm] \mathcal{O}_i\end{array}\hspace{-1mm}}

\newcommand{\OBW}{\mathcal{O}_{\phi\widetilde{W}B}}
\newcommand{\JT}[1]{\color{red}{ JT: \bf #1}\color{black}}
\usepackage{longtable}
\usepackage{float}
\usepackage{wrapfig}
\usepackage{tikz}
\usetikzlibrary{calc,trees,positioning,arrows,chains,shapes.geometric,%
    decorations.pathreplacing,decorations.pathmorphing,shapes,%
    matrix,shapes.symbols}
\usepackage{tikz-feynman,contour}
\tikzfeynmanset{compat=1.1.0}
\tikzfeynmanset{warn luatex=false}
\usepackage{pstricks}
\usepackage{soul}
\usepackage{textcomp}
\usepackage{marvosym}
\usepackage{wasysym}
\usepackage{amsmath,amssymb}
\usepackage{cancel}
\usepackage{bbold}
\usepackage{slashed}
\usepackage{latexsym}
\usepackage{hyperref}
\usepackage{cite}
\usepackage[english]{babel}
\usepackage[normalem]{ulem}
\usepackage{braket}
\usepackage{breqn}
\usepackage{mathrsfs}
\usepackage{mathtools}
\usepackage[most]{tcolorbox} 
\usepackage{booktabs}
\usepackage{multirow}
\usepackage[absolute,overlay]{textpos}
\usepackage{xmpmulti} % multi image include
\usepackage{smartdiagram}
\usepackage{metalogo}
\usetikzlibrary{calc,positioning,shadows.blur,decorations.pathreplacing}
\usepackage{etoolbox}

\title{A Reduced basis for CP violation in SMEFT at colliders and its application to Diboson production}
\author{Céline Degrande, Julien Touchèque}
\date{\today}

\begin{document}

\maketitle
\thispagestyle{empty}

\begin{abstract}
\begin{comment}
\JT{Propositions for new title:}
\begin{itemize}
    \item A Reduced Bases for CP violation in SMEFT and Sensible Observables in Diboson production
    \item CP violation in SMEFT : Reduced Bases, CP-sensitive observables and Diboson production
    \item Reduced SMEFT bases for leading CP-violation effects at colliders and its application to find efficient observable in diboson production
    \item A Reduced basis for CP violation in SMEFT
    %, CP-sensitive observables (
    at colliders 
    %/LHC) 
    and its application to Diboson production
    \item Leading CP-violating operators in SMEFT and Sensitive observables : Example in diboson production
\end{itemize}
\end{comment}
We show that only 10 (17) CP-odd operators of the SMEFT give the leading, \textit{i.e.} least suppressed by the new physics scale, CP-violating contributions once we assume that all fermions are massless but the top (and bottom) quark(s). We start with a short review of previous analyses focusing on operators of our reduced basis and list different observables probing their CP violating effects by direct measurements at colliders and by indirect measurements in low-energy observables. Since CP-odd operators typically lead to phase space suppressed interferences, we quantify the efficiency to revive the interference for various observables found in the literature but also for new observables in diboson production. Our new observables are found to be more efficient on the whole experimental fiducial phase space and are complementary to those presented so far as they probe different combinations of operators and get their sensitivities from different regions of the phase space.
\end{abstract}

\cleardoublepage
\setcounter{page}{1}
\newpage

\tableofcontents

\section{Introduction}
\input{intro.tex}

\section{Reduced basis and motivations}\label{sec:theory}

\subsection{Leading CP contributions in SMEFT}
\input{smeft}

\subsection{Basis Reduction : $U(1)^{14}$ symmetry}\label{subsec:basis reduction u14}
\input{basisred_1}

\subsection{Basis Reduction : $U(1)^{13}$ symmetry}\label{subsec:basis reduction u13}
\input{basisred_2}

\subsection{Sign of the interference}\label{subsec:signofinterference}
\input{signandPOI}

\subsection{Review of CP-sensible observables}\label{subsec:review}
\input{review}

\section{Analysis}\label{sec:analysis}

\input{analysis}

\section{Results}\label{sec:results}
\input{results}

\section{Discussion and Conclusion}\label{sec:conclusion}
\input{conclusion}

\bibliographystyle{unsrt}
\bibliography{bibliography}

\end{document}

%% file: intro.tex
The discovery of the long-awaited Higgs boson \cite{Aad:2012tfa,Chatrchyan:2012xdj} as well as the vast majority of LHC measurements conducted so far largely corroborate the Standard Model (SM) predictions \cite{Giardino:2013bma,Corbett:2015ksa,Khachatryan:2016vau,Diaz-Cruz:2019vka,Yao:2019lrt,Sopczak:2020vrs}. However, the SM is not flawless. In particular, the SM fails to explain the matter-antimatter asymmetry in the Universe. The theory of electroweak baryogenesis is a strong candidate to solve the matter-antimatter  mystery. While the SM CKM phase breaks CP as requested by the Sakharov's conditions \cite{Sakharov:1967dj}, its effects are not large enough and therefore additional sources of CP-violation (CPV) are required (see Ref.~\cite{Canetti:2012zc} for a review). 

As a result, it seems that the SM is valid only at the energies probed so far and should be completed by New Physics (NP) at higher energies. Therefore, we use the effective field theory extension of the SM (SMEFT) \cite{Buchmuller:1985jz, Grzadkowski:2010es} as a model-independent way to parameterize the effects of heavy NP. In this framework, any extra CP violating sources introduced by the heavy NP can be expressed in terms of CP violating SMEFT parameters. In particular, we use the CP-odd operators of the SMEFT not only to identify the most promising processes but also to design observables such that they offer a good sensitivity to those CP-odd operators. We look for simple observables which do not heavily rely on the details of the SMEFT or on the expression of the amplitude. Therefore, they could be easily reinterpreted if the SMEFT assumption turns out to be disproved by the data. In addition, the measurements of those simple observables are easier to perform than optimised measurements based on methods such as the matrix element method \cite{Brehmer:2018kdj, Brehmer:2018eca}. Our hope is to find simple observables with a sensitivity close enough to those of advanced methods to indicate the presence or absence of NP and thus to point out where to apply those more expensive analyses. To this end, we use the method developed in Ref.~\cite{Degrande:2020tno} to quantify the efficiency of those simple observables to revive the interference between the CP-odd operators and the SM.
 
This paper is outlined as follows. We trim the list of CP violating operators to keep only those that can have large effects at colliders and give a review of the literature on those operators in Section \ref{sec:theory}. Section \ref{sec:analysis} focuses on CP violating effects in diboson production. We investigate the $WZ$ and $W\gamma$ in their leptonic channels and we display the results in section \ref{sec:results}. We then compare them to other observables already developed in \cite{Azatov:2019xxn,Kumar:2008ng,Dawson:2013owa}. We summarize and conclude in Section \ref{sec:conclusion}. 

%% file: smeft.tex
So far no significant deviation from the SM has been observed at colliders and no new resonances have been found at the LHC beyond the Higgs boson. On the contrary, bounds on new resonances and NP are getting stronger. This observation can be interpreted as a hint that a significant energy gap exists between the LHC reach and the NP scale $\Lambda$. If $\Lambda$ sits sufficiently above the TeV range then NP naturally decouple from the SM \cite{Appelquist:1974tg}. Therefore, we can parameterize all NP effects with an effective theory at energies $E<<\Lambda$. In this work, we assume that the symmetries of the SM are preserved by higher order operators \textit{i.e.}, we use the framework of the SMEFT. The SMEFT Lagrangian, $\mathcal{L}_{SMEFT}$, is defined as
\begin{equation}\label{smeft expansion}
    \mathcal{L}_{SMEFT} = \mathcal{L}_{SM} + \sum_{d=5}^{\infty} \frac{1}{\Lambda^{d-4}} \mathcal{L}_{d},
\end{equation}
where the first term $\mathcal{L}_{SM}$ corresponds to the usual SM Lagrangian and the $\mathcal{L}_{d}$ terms contain operators of dimension $d$, $\{\mathcal{O}^d_i\}_{i,d>4}$ with their Wilson coefficients $\{C_i\}_i$, parameterizing all the new interactions originating from the unknown UV-complete theory. 

The infinite number of coefficients does not allow a complete analysis therefore only the leading terms of the $1/\Lambda$ expansion are kept to focus on the largest NP contributions. 
%\CD{check spelling UK or US everywhere}
%
At the Lagrangian level, as each $\mathcal{O}^d_i$ is suppressed by $d-4$ powers of $\Lambda$, the first NP contribution sits at the $\Lambda^{-1}$ order and consists in the single Weinberg operator $\mathcal{L}_5$. This term provides a neutrino mass term and violates the conservation of the lepton number $L$. However, we assume here that $L$ is conserved and therefore we neglect $\mathcal{L}_5$ leaving the $\Lambda^{-2}$ order as the leading source of NP effects. As a matter of fact, we truncate the Lagrangian expansion at the dimension-6 terms $\mathcal{L}_6$,
\begin{equation}\label{operator expansion}
    \mathcal{L}_{SMEFT} \sim \mathcal{L}_{SM} + \frac{1}{\Lambda^{2}} \mathcal{L}_{6} = \mathcal{L}_{SM} + \sum_{i} \frac{C_i}{\Lambda^{2}} \mathcal{O}_i + \sum_{i} \Big( \frac{\hat{C}_i}{\Lambda^{2}} \hat{\mathcal{O}}_i + h.c. \Big).
\end{equation}
The operators in $\mathcal{L}_6$ are classified in two categories. The $\mathcal{O}_i$ category represents hermitian operators, whereas the $\hat{\mathcal{O}}_i$ category represents non-hermitian operators. 
The set $\left\{ \mathcal{O}_i , \hat{\mathcal{O}}_{i'} \right\}$ acts as a basis in the dimension-six operator space with their respective Wilson coefficients $\left\{C_i , \hat{C}_i \right\}$. 
In the absence of absorptive phases\footnote{Absorptive complex phases can arise when new degrees of freedom at low energies create the dimension-six operators as discussed in Ref.\cite{Brehmer:2017lrt}. This case is not considered in this paper. }, $C_i$'s must be real by hermiticity. On the contrary, $\hat{C}_i$'s can be complex which is an important distinction when considering CP-odd operators. 
Operators originating from the $\mathcal{O}_i$ class immediately introduce CP violating effects while operators from the $\hat{\mathcal{O}}_i$ class  rely on the imaginary part of their coefficients to create such effects. 
Most of the time, global fits (see Refs. \cite{Buckley:2015nca,Ellis:2018gqa,Hartland:2019bjb,Brivio:2019ius,Falkowski:2019hvp,Basan:2020btr,Bissmann:2020mfi,Bissmann:2019gfc} for some examples of such analyses) include non-hermitian operators but overlook CP violating effects in their analysis by imposing CP symmetry, \textit{i.e.} $\hat{C}_i$'s are considered as real and the hermitian CP-odd operators are discarded.

%full dim6 operator table
\begin{table}
\centering
\scalebox{0.95}{%
\begin{tabular}{|c|c|c|c|c|c|}
\hline 
   \multicolumn{2}{|c|}{$(X^3)$}  & \multicolumn{2}{|c|}{$(\psi^2\phi^3)$}  &  \multicolumn{2}{|c|}{$(\psi^2\phi^2 D)$}  \\
\hline
  $O_{\Tilde{G}GG}$ & $f^{ABC}\widetilde{G}_\mu^{A\nu} G_\nu^{B\rho} G_\rho^{C\mu}$  &  $O_{u\phi}$ & $(\phi^\dagger\phi)(\overline{q} u \Tilde{\phi})$ & $O_{\phi u d}$ & $i(\Tilde{\phi}^\dagger D_\mu\phi)(\overline{u} \gamma^\mu d)$  \\
  $O_{\Tilde{W}WW}$ & $\epsilon^{IJK}\widetilde{W}_\mu^{I\nu} W_\nu^{J\rho} W_\rho^{K\mu}$ & $O_{d\phi}$ & $(\phi^\dagger\phi)(\overline{q} d \phi)$ & & \\
   & & $O_{e\phi}$ & $(\phi^\dagger\phi)(\overline{l} e \phi)$ & & \\
\hline \hline 
  \multicolumn{2}{|c|}{$(X^2\phi^2)$} & \multicolumn{2}{|c|}{$(\psi^4)$} & \multicolumn{2}{|c|}{$(X\psi^2\phi)$} \\
\hline
  $O_{\phi \Tilde{G}}$ & $\phi^\dagger\phi\widetilde{G}_{\mu\nu}^A G^{A\mu\nu}$ & $O_{ledq}$ & $(\overline{l}^j e)(\overline{d} q^j)$ & $O_{uG}$ & $(\overline{q} \sigma^{\mu\nu}T^A u)\Tilde{\phi}G^A_{\mu\nu}$  \\
  $O_{\phi \Tilde{W}}$ & $\phi^\dagger\phi\widetilde{W}^I_{\mu\nu} W^{I\mu\nu}$ & $O_{lequ}^{(1)}$ & $(\overline{l}^j e)\epsilon_{jk}(\overline{q}^k u)$ & $O_{uW}$ & $(\overline{q} \sigma^{\mu\nu}u)\tau^I\Tilde{\phi}W^I_{\mu\nu}$  \\
  $O_{\phi \Tilde{B}}$ & $\phi^\dagger\phi\widetilde{B}_{\mu\nu} B^{\mu\nu}$  & $O_{lequ}^{(3)}$ & $(\overline{l}^j \sigma^{\mu\nu} e)\epsilon_{jk}(\overline{q}^k \sigma_{\mu\nu} u)$  & $O_{uB}$ & $(\overline{q}\sigma^{\mu\nu}u)\Tilde{\phi}B_{\mu\nu}$ \\
  $O_{\phi \Tilde{W}B}$ & $\phi^\dagger\tau^I\phi\widetilde{W}^I_{\mu\nu} B^{\mu\nu}$  & $O_{quqd}^{(1)}$ & $(\overline{q}^j u)\epsilon_{jk}(\overline{q}^k d)$ & $O_{dG}$ & $(\overline{q}\sigma^{\mu\nu}T^A d)\phi G^A_{\mu\nu}$  \\
    &  & $O_{quqd}^{(8)}$ & $(\overline{q}^j T^A u)\epsilon_{jk}(\overline{q}^k T^A d)$ & $O_{dW}$ & $(\overline{q} \sigma^{\mu\nu} d)\tau^I\phi W^I_{\mu\nu}$ \\
    &  & & & $O_{dB}$ & $(\overline{q} \sigma^{\mu\nu} d)\phi B_{\mu\nu}$ \\
   & & & & $O_{eW}$ & $(\overline{l} \sigma^{\mu\nu} e )\tau^I\phi W^I_{\mu\nu}$ \\
   & & & & $O_{eB}$ & $(\overline{l} \sigma^{\mu\nu} e)\phi B_{\mu\nu}$ \\
\hline
\end{tabular}}
\caption{List of CP-odd dimension-6 operators present in the Warsaw basis for one fermion generation. }
\label{complete CPV operator basis}
\end{table}

Before presenting the relevant SMEFT operators, we first detail our notation. The field $\psi$ represents the fermionic fields present in the SM: the left-handed lepton doublets $l_L$, right-handed charged leptons $e_R$, left-handed quark doublets $q_L$, right-handed up-quarks $u_R$ and right-handed down-quarks $d_R$. The chirality indices $L$ and $R$ are deemed implicit in the following. The gauge field $X$ stands for any gauge field strength tensors $G^A, W^I, B$ and dual tensors are defined by
\begin{equation*}
    \widetilde{X}_{\mu\nu}=\frac{1}{2}\epsilon_{\mu\nu\rho\sigma} X^{\rho\sigma},
\end{equation*}
where $\epsilon_{\mu\nu\rho\sigma}$ is the totally antisymmetric tensor ($\epsilon_{0123}=+1$). The indices $\{I,J,K\}$ and $\{A,B,C\}$ are respectively isospin and colour indices. The scalar $\phi$ is the usual  $SU(2)_L$ doublet and $\widetilde{\phi}$ is its conjugate defined by $\widetilde{\phi}^i = \epsilon_{ij} (\phi^{*})^j$ ($\epsilon_{12}=+1$). We conform our notation to the Warsaw basis~\cite{Grzadkowski:2010es} by normalizing the $SU(3)_c$ and $SU(2)_L$ generators as $Tr(T^AT^B)=\frac{1}{2}\delta^{AB}$ and $Tr(\tau^I\tau^J)=2\delta^{IJ}$. 

We choose to represent dimension-six operators in the non-redundant Warsaw basis~\cite{Grzadkowski:2010es}. The operators have been counted in Appendix A of Ref.~\cite{Alonso:2013hga}: 23 CP-odd operators for one fermion generation and 1349 for three fermion generations. Table~\ref{complete CPV operator basis} lists the operators in the unbroken phase for one fermion generation and sorts them in 6 classes according to their particle contents. We have not represented the CP-odd operators for three fermion generations for a reason that will be mentioned later. 

The left column of Table~\ref{complete CPV operator basis} is composed of the pure gauge boson $X^3$ and the Higgs-gauge boson $X^2\phi^2$ classes. %They respectively contain the hermitian operators $\{\mathcal{O}_{\Tilde{G}GG}, \mathcal{O}_{\Tilde{W}WW} \}$ and $\{ \mathcal{O}_{\phi\Tilde{G}}, \mathcal{O}_{\phi\Tilde{W}}, \mathcal{O}_{\phi\Tilde{B}}, \mathcal{O}_{\phi\Tilde{W}B} \}$; 
These operators belong to the $\mathcal{O}_i$ category. The 4-fermion $\psi^4$ and the scalar-fermion $\psi^2 \phi^3$ classes are presented in the central column. For one generation only 5 operators are present in the $\psi^4$ class: one $(\Bar{L}R)(\Bar{R}L)$ operator %$\{\mathcal{O}_{ledq}\}$ 
and four $(\Bar{L}R)(\Bar{L}R)$ operators %$\{  \mathcal{O}_{lequ}^{(1)}, \mathcal{O}_{lequ}^{(3)}, \mathcal{O}_{quqd}^{(1)}, \mathcal{O}_{quqd}^{(8)} \}$
. The $\psi^2 \phi^3$ class includes three operators that modify the interactions between the Higgs and the fermions%$\{ \mathcal{O}_{u\phi}, \mathcal{O}_{d\phi}, \mathcal{O}_{e\phi} \}$
. Finally, the right column contains the dipole class $X\psi^2\phi$ with its eight operators %$\{ \mathcal{O}_{uG}, \mathcal{O}_{uW}, \mathcal{O}_{uB}, \mathcal{O}_{dG}, \mathcal{O}_{dW}, \mathcal{O}_{dB}, \mathcal{O}_{eW}, \mathcal{O}_{eB} \}$ and the operator $\{\mathcal{O}_{\phi u d }\}$ from $\psi^2 \phi^2 D$
. The operators from the central and right columns are not hermitian thus belong to the $\hat{\mathcal{O}}_i$ category. The Wilson coefficients
\begin{equation}
    \{ C_{\Tilde{G}GG}, C_{\Tilde{W}WW}, C_{\phi\Tilde{G}}, C_{\phi\Tilde{W}}, C_{\phi\Tilde{B}}, C_{\phi\Tilde{W}B} \}
\end{equation}
are real and CP is broken when they are non-vanishing, while, for one fermion generation,
\begin{multline}
    \{ C_{u\phi}, C_{d\phi}, C_{e\phi}, C_{\phi u d}, C_{ledq}, C_{lequ}^{(1)}, C_{lequ}^{(3)}, C_{quqd}^{(1)},C_{quqd}^{(8)}, \\ C_{uG}, C_{uW}, C_{uB}, C_{dG}, C_{dW}, C_{dB}, C_{eW}, C_{eB} \}
\end{multline}
are complex numbers and their imaginary parts induce new sources of CPV.

However, one comment needs to be addressed about CP-odd operators from both $\psi^4$ and $\psi^2\phi^2 D$ classes. The $\psi^4$ class is the most sensitive to the number of generations and is by far the largest class in terms of CP-odd operators. As soon as more generations are taken into consideration, the number of operators skyrockets with up to 1014 CP-odd operators for three generations. The reason is that the central column of Table \ref{complete CPV operator basis} would include additional 4-fermions operators with different chirality configurations, \textit{i.e.} it would also contain $(\Bar{L}L)(\Bar{L}L)$, $(\Bar{R}R)(\Bar{R}R)$ and $(\Bar{L}L)(\Bar{R}R)$ operators if these operators mix fermions of different generations. For example, if the two fermion fields in one fermion pair are from different generations, the operator is no longer hermitian: the operator $(\Bar{e}\gamma_\mu e)(\Bar{e}\gamma^\mu e)$ is hermitian and therefore CP-even while the operator with a substitution of an electron by a muon, $(\Bar{\mu}\gamma_\mu e)(\Bar{e}\gamma^\mu e)$, is non-hermitian and can induce CPV if its Wilson coefficient has an imaginary part. The argument of the generation indices also stands in the $\psi^2\phi^2 D$ class and provides additional CP-odd operators, up to 30 for three generations instead of 1 for one generation. The complete list of 4-fermion and $\psi^2\phi^2 D$ operators is given in Ref.~\cite{Grzadkowski:2010es}.

In this formalism, the total amplitude is
\begin{equation}
\mathcal{M}=\mathcal{M}_{SM} + \sum_i \mathcal{M}_{i} + \mathcal{O}\left(\Lambda^{-4}\right)\qquad
\end{equation} 
where $\mathcal{M}_{SM}$ is the SM amplitude and $\mathcal{M}_i$ the amplitude involving one vertex from one relevant CP-odd dimension-six operator either $\mathcal{O}_i$ or $\hat{\mathcal{O}}_i$. 
%\CD{changed the notation to make sure we are not excluding the hat operators, tried $\obothi$ but too big for the subscript of the amplitude}
%\CD{can we put a hat between brackets  %$\overset{\scriptscriptstyle{\left({\wedge}\right)}}{\mathcal{O}}$ or
%\vspace{-2mm}$\begin{array}{c}\scriptscriptstyle{\left({\wedge}\right)\,}\\[-1.7mm] \mathcal{O}_i\end{array}$
%$\obothi$ or $\hspace{-1mm}\begin{array}{c}\quad\\[-8mm]\scriptscriptstyle{\left({\wedge}\right)\,}\\[-1.7mm] \mathcal{O}_i\end{array}\hspace{-1mm}$}.
%
Squaring the total amplitude gives 
\begin{equation}\label{full amplitude expansion}
     \| \mathcal{M} \|^2  = \| \mathcal{M}_{SM} \|^2 + \sum_i \mathcal{M}_{int,i} + \sum_i \| \mathcal{M}_{i} \|^2 + \sum_{i\neq j} \mathcal{M}_{int,ij}+ \mathcal{O}\left(\Lambda^{-4}\right).
     %&  \| \mathcal{M}_{SM} \|^2 + \sum_i 2 \Re e  \Big\{ \mathcal{M}_{SM}^* \mathcal{M}_{\mathcal{O}_i} \Big\} + \sum_i \| \mathcal{M}_{\mathcal{O}_i} \|^2 + \sum_{i\neq j} 2 \Re e  \Big\{ \mathcal{M}_{\mathcal{O}_i}^* \mathcal{M}_{\mathcal{O}_j} \Big\}+ \mathcal{O}\left(\Lambda^{-4}\rigth), \nonumber \\
        %& =
\end{equation}
where $\mathcal{M}_{int,i} \equiv 2 \Re e  \Big\{ \mathcal{M}_{SM}^* \mathcal{M}_{i} \Big\}$ and $\mathcal{M}_{int,ij} \equiv 2 \Re e  \Big\{ \mathcal{M}_{i}^* \mathcal{M}_{j} \Big\}$. 
We neglect CPV from $\mathcal{M}_{SM}$ in the first term of Eq.(\ref{full amplitude expansion}) as CP violating effects from the CKM phase $\delta_{CKM}$ are negligible\footnote{This is demonstrated by the small value of the Jarkslog invariant generated by $\delta_{CKM}$ \cite{Jarlskog:1985ht,PartDataGroup}.}. 
Just like with the operators, we will consistently neglect any $\Lambda^{-4}$ contributions in our analysis to focus on the leading contributions from the CP-odd operators, embodied by the $\Lambda^{-2}$ terms. In Eq.(\ref{full amplitude expansion}), they come from the interference between $\mathcal{M}_{SM}$ and $\mathcal{M}_{i}$. The third and fourth terms are $\Lambda^{-4}$-suppressed so they are discarded. 
In fact, obtaining the full $\Lambda^{-4}$ contributions would require to include dimension-eight operators and to consider diagrams with two vertices from dimension-six operators which, by interfering with the SM amplitude, contribute at that order. 
As a result, we can study each dimension-six operator in the amplitude independently since the NP contribution is only linear in the Wilson coefficients,
\begin{equation}\label{amplitude expansion}
   \| \mathcal{M} \|^2 \sim  \| \mathcal{M}_{SM} \|^2 + \sum_i \mathcal{M}_{int,i}.
\end{equation}
%In the end, we are still left with 1149 degrees of freedom associated with the Wilson coefficients of the Warsaw basis without indications on which operators create the leading CP violating contributions. 

%% file: basisred_1.tex
Dimensional analysis tells us that the leading energy dependence in $\mathcal{M}_{int}$ from Eq.(\ref{amplitude expansion}) can have a growth factor $\frac{E^2}{\Lambda^{2}}$ compared to $\mathcal{M}_{SM}$ where $E$ is the/one of the energy of the process. This energy growth is expected in amplitudes from non-renormalizable operators and begins to be more significant when $\frac{E^2}{\Lambda^{2}} \lesssim 1$, that is $E \lesssim \Lambda$. Comparatively, contributions proportional to fermion masses $m_f$ become negligible at high energy. Since our goal is to track the leading CP violating contributions in high energy colliders, we neglect all the fermion masses but the top quark mass which we consider too close to the TeV range to be discarded.

To compel our analysis to this hypothesis, we only need to impose $U(1)^{14}$ at high energies \footnote{This symmetry has less generators than the $U(3)_l \times U(3)_e \times U(3)_d \times U(2)_q \times U(2)_u$ symmetry often used in LHC analyses to restrict the analysis to the top sector.}. The Yukawa terms in $\mathcal{L}_{SM}$ violate the new symmetry so all Yukawa couplings $\Gamma_f=m_f/v$, for $f=\{q,l,u,d,e\}$, must vanish except for the top quark, $\Gamma_t\neq 0$,
\begin{equation}
    \mathcal{L}_{SM} \supset \Bar{e} \Gamma_l \phi l + \Bar{d} \Gamma_d \phi q + \Bar{u} \Gamma_u \widetilde{\phi} q \xrightarrow
 {\substack{U(1)^{14}}} \Bar{t} \Gamma_t \widetilde{\phi} t .
\end{equation}
As a result, only the four heaviest SM particles remain massive after the electroweak symmetry breaking (EWB): the top quark $t$, the Higgs boson $h$ and the weak vector bosons $W^{\pm}, Z$. 
Moreover, left-handed and right-handed light fermion fields also decouple since their coupling through the Higgs boson in $\mathcal{L}_{SM}$ is gone. This is particularly interesting because we are now allowed to choose the phase of each fermion field independently. 
We rephase light fermion fields in dimension-six operators of $\mathcal{L}_{SMEFT}$ such that phases of Wilson coefficients are absorbed. We take as an example the $\mathcal{O}_{bG}$ operator from the dipole class and its Wilson coefficient $C_{bG}$. This operator introduces a new vertices contributing to $pp \rightarrow b\Bar{b}h$. Let us define the phase $\varphi_i$ of the Wilson coefficient $C_i$, $C_i \equiv e^{i\varphi_i}|C_i|$. We rephase the right-handed bottom field $b_R$ with the opposite phase of $C_{bG}$, namely
\begin{equation}\label{eq:brephasing}
    b_R \rightarrow e^{-i \varphi_{bG}} b'_R,
\end{equation}
such that $\mathcal{L}_{SM}(m_f\rightarrow 0)$ is unaffected but the $\mathcal{O}_{bG}$ operator together with its Wilson coefficient becomes
\begin{equation}\label{eq:ObGrephasing}
   e^{i\varphi_{bG}} |C_{bG}| (\Bar{Q} \sigma^{\mu\nu} T^A b) \Tilde{\phi} G^A_{\mu\nu}  \rightarrow |C_{bG}| (\Bar{Q} \sigma^{\mu\nu} T^A b') \Tilde{\phi} G^A_{\mu\nu}  .
\end{equation}
Due to its real Wilson coefficient, when added to its self-conjugate in $\mathcal{L}_{SMEFT}$, the $\mathcal{O}_{bG}$ operator no longer gives a CP violating contribution at $\Lambda^{-2}$ order. We can apply the rephasing to all operators one at a time and reduce the list of CP-odd operators from Table \ref{complete CPV operator basis}. 
We are allowed to do so because we are restricting ourselves $\Lambda^{-2}$ corrections, \textit{i.e.} the interferences between the SM and one CP-odd operator, which are linear in the Wilson coefficients. This simplification would not stand if we were looking at $\Lambda^{-4}$ corrections. 

As stated above, at the $\Lambda^{-4}$ order, the amplitude can simultaneously depend on two Wilson coefficients in the interference between 2 dimension-six operators. In this case, the extra phase cannot be absorbed and will be transferred from one operator to other instead. For instance, still in $pp \rightarrow b\Bar{b}h$, $\mathcal{O}_{bG}$ and $\mathcal{O}_{b\phi}$  are two relevant operators and, under the transformation outlined in Eq.(\ref{eq:brephasing}), they become
\begin{eqnarray}
   && e^{i\varphi_{bG}} |C_{bG}| (\Bar{Q} \sigma^{\mu\nu} T^A b) \Tilde{\phi} G^A_{\mu\nu}  \rightarrow |C_{bG}| (\Bar{Q} \sigma^{\mu\nu} T^A b') \Tilde{\phi} G^A_{\mu\nu},  \nonumber \\
   && e^{i\varphi_{b\phi}} |C_{b\phi}| (\Bar{Q}  b\phi)   \left(\phi^\dagger \phi\right)  \rightarrow e^{i(\varphi_{b\phi}-\varphi_{bG})} |C_{b\phi}| (\Bar{Q}  b'\phi)   \left(\phi^\dagger \phi\right).
   \label{eq:bop}
\end{eqnarray}
We still obtain a real $C_{bG}$ as in (\ref{eq:ObGrephasing}) but this time its phase is passed to $C_{b\phi}$, therefore transferring the CP violating effects. 

Sticking to the $\Lambda^{-2}$ order, we use the rephasing trick independently to every operator containing light fermion fields. %Therefore, CP-odd operators remaining after the rephasing are composed of massive fields e.g. the top quark, the Higgs boson and the electroweak bosons $W$, $Z$. 
The massless approximation does not affect the hermitian operators in the bosonic classes $X^3$ and $X^2\phi^2$, and they remain present in the reduced basis: 
\begin{equation*}
    \{ \mathcal{O}_{\widetilde{G}GG}, \mathcal{O}_{\widetilde{W}WW}, \mathcal{O}_{\phi \widetilde{G}}, \mathcal{O}_{\phi \widetilde{W}}, \mathcal{O}_{\phi \widetilde{B}}, \mathcal{O}_{\phi \widetilde{W}B} \}.
\end{equation*} 
In fact, those operators are even invariant under the full $U(1)^{15}$ symmetry, which remove all the other CP odd operators, as well as under the full flavour symmetry of the SM when all the fermions are massless, $U(3)^5$~ \cite{Gerard:1982mm}. However, the simplification drastically decreases the number of operators in the four other classes because the operators contain light fermion fields or the  right-handed bottom field. Operators with fermion fields remain only if they consist in bosons with left- and right-handed top quark fields only. The $\psi^4$ and $\psi^2\phi^2 D$ classes become free of CP-odd operators, $\mathcal{O}_{t\phi}$ is the only remaining operator in the $\psi^2\phi^3$ class and one operator for each vector boson field associated to top quark fields remain in the $X \psi^2\phi^2$ class. Therefore, we keep in our basis :
\begin{equation*}
    \{ \mathcal{O}_{tG}, \mathcal{O}_{tW}, \mathcal{O}_{tB}, \mathcal{O}_{t \phi} \}.
\end{equation*}
In the end, 10 CP-odd operators remain under $U(1)^{14}$ and they are listed in Table~\ref{New CPV operator basis u14} following the class ordering from Table~\ref{complete CPV operator basis}. 

\begin{table}[t]
\centering
\begin{tabular}{|c|c|c|c|c|c|}
\hline 
   \multicolumn{2}{|c|}{$(X^3)$}  & \multicolumn{2}{|c|}{$(\psi^2\phi^3)$}  &  \multicolumn{2}{|c|}{$(\psi^2\phi^2 D)$}  \\
\hline
  $O_{\Tilde{G}GG}$ & $f^{ABC}\widetilde{G}_\mu^{A\nu} G_\nu^{B\rho} G_\rho^{C\mu}$  &  $O_{t\phi}$ & $(\phi^\dagger\phi)(\overline{q}_r t_r\Tilde{\phi})$ & // & /////  \\
  $O_{\Tilde{W}WW}$ & $\epsilon^{IJK}\widetilde{W}_\mu^{I\nu} W_\nu^{J\rho} W_\rho^{K\mu}$ &  &  & & \\
\hline \hline 
  \multicolumn{2}{|c|}{$(X^2\phi^2)$} & \multicolumn{2}{|c|}{$(\psi^4)$} & \multicolumn{2}{|c|}{$(X\psi^2\phi)$} \\
\hline
  $O_{\phi \Tilde{G}}$ & $\phi^\dagger\phi\widetilde{G}_{\mu\nu}^A G^{A\mu\nu}$ & // & ///// & $O_{tG}$ & $(\overline{q}_3 \sigma^{\mu\nu}T^A t)\Tilde{\phi}G^A_{\mu\nu}$  \\
  $O_{\phi \Tilde{W}}$ & $\phi^\dagger\phi\widetilde{W}^I_{\mu\nu} W^{I\mu\nu}$ &  &  & $O_{tW}$ & $(\overline{q}_3 \sigma^{\mu\nu}t)\tau^I\Tilde{\phi}W^I_{\mu\nu}$  \\
  $O_{\phi \Tilde{B}}$ & $\phi^\dagger\phi\widetilde{B}_{\mu\nu} B^{\mu\nu}$  &  &   & $O_{tB}$ & $(\overline{q}_3 \sigma^{\mu\nu}t)\Tilde{\phi}B_{\mu\nu}$ \\
  $O_{\phi \Tilde{W}B}$ & $\phi^\dagger\tau^I\phi\widetilde{W}^I_{\mu\nu} B^{\mu\nu}$  &  &  &  &  \\
\hline
\end{tabular}
\caption{List of CP-odd dimension-6 operators in our reduced basis under $U(1)^{14}$. The operators of the first column are invariant under the full $U(3)^5$ flavour symmetry. }
\label{New CPV operator basis u14}
\end{table}

It is important to keep in mind that the massless fermions approximation does not affect the real part of Wilson coefficients, \textit{i.e.} the CP-even operators and the CP-even part of non-hermitian operators. Thus, a complete reduced basis at the $\Lambda^{-2}$ order would not only contain the 10 CP-odd operators from Table \ref{New CPV operator basis u14} but all CP-even operators with 3 generations from Table \ref{complete CPV operator basis} invariant under the chosen symmetry as well.

The leading CP violating contributions we were looking for come from these 10 operators which not only introduce anomalous couplings with respect to the SM but introduce new contact interactions as well, as showed in Table \ref{tab:couplings}.

\begin{table}[]
    \centering
    \begin{tabular}{|c|c|}
        \hline
        Operator & Vertices \\
        \hline
        $\mathcal{O}_{\widetilde{G}GG}$ & ggg/gggg/ggggg/gggggg \\
        \hline
        $\mathcal{O}_{\widetilde{W}WW}$ & WWB/WWBB/WWWW/WWBBB/WWWWB \\
        \hline
        $\mathcal{O}_{\phi \widetilde{G}}$ & hgg/hggg/hgggg/hhgg/hhggg/hhgggg \\
        \hline
        $\mathcal{O}_{\phi \widetilde{W}}$ & hBB/hhBB/hWW/hhWW/hhBWW \\
        \hline
        $\mathcal{O}_{\phi \widetilde{B}}$ & hBB/hhBB \\
        \hline
        $\mathcal{O}_{\phi \widetilde{W}B}$ & WWB/WWBh/WWBhh/BBh/BBhh  \\
        \hline
        $\mathcal{O}_{t \phi}$ & tth/tthh/tthhhh  \\
        \hline
        $\mathcal{O}_{t G}$ & ttg/ttgh/ttgg/ttggh  \\
        \hline
        $\mathcal{O}_{t W}$ & ttB/ttBh/ttWW/ttWWh/tbW/tbWh/tbWBh  \\
        \hline
        $\mathcal{O}_{t B}$ & ttB/ttBh  \\
        \hline
    \end{tabular}
    \caption{Interactions introduced by the operators in Table \ref{New CPV operator basis u14} in the unitary gauge. W stands for the charged gauge boson, B for the neutral electroweak boson, \textit{ i.e.} the Z boson and the photon, h for the higgs, g for the gluon, t for the top quark and b for the bottom quark. %No \JT{Do the same for new operators with $U(1)^{13}$ ?}
    }
    \label{tab:couplings}
\end{table}

%% file: basisred_2.tex
The $U(1)^{14}$ symmetry introduced in the previous section is not a subgroup of the $U(3)_l \times U(3)_e \times U(3)_d \times U(2)_q \times U(2)_u$ symmetry often used for SMEFT analyses related to the top sector contrary to $U(1)^{13}$. In this latter symmetry, both top and bottom quarks are massive. Therefore, there is less freedom to cancel the phases of CP-odd operators. For example, the phase of the operators in Eq.\eqref{eq:bop} cannot be removed anymore.

Therefore, the operators present under the $U(1)^{14}$ symmetry must remain and are supplemented by other operators containing bottom quark fields. As already mentioned, the operators from $X^3$ and $X^2\phi^2$ classes remain unaffected by any redefinition as they are invariant under the $U(3)^5$ symmetry.
By following the same method than Subsection \ref{subsec:basis reduction u14}, the non-hermitian operators of the $\psi^2\phi^3$ and $X\psi^2\phi$ categories including the right bottom quark fields are:
\begin{equation*}
    \{ \mathcal{O}_{bG}, \mathcal{O}_{bW}, \mathcal{O}_{bB}, \mathcal{O}_{b \phi} \}.
\end{equation*}

However, this symmetry also allows operators of the two other categories. We mentioned previously that $O_{\phi ud}$ was the only CP-odd operator in the $\psi^2 \phi^2 D$ and, using the up and down right-handed fields, we now need to include :
\begin{equation*}
    \{ \mathcal{O}_{\phi tb}  \}.
\end{equation*}
To build non-hermitian four fermions operators with chirality configurations $(\Bar{L}L)(\Bar{L}L)$, $(\Bar{R}R)(\Bar{R}R)$ and $(\Bar{L}L)(\Bar{R}R)$, we need fields from different generations. Since we are dealing only with the third generation, no operators come from these configurations. The $(\Bar{L}R)(\Bar{R}L)$ configuration involves only leptons so we do not have CP-odd operators from those either. Finally, $(\Bar{L}R)(\Bar{L}R)$ has two operators without leptons which can be CP-odd : 
\begin{equation*}
    \{ \mathcal{O}_{qtqb}^{(1)}, \mathcal{O}_{qtqb}^{(8)}  \}.
\end{equation*}

Here, 17 CP-odd operators remain under $U(1)^{13}$ and they are listed in Table~\ref{New CPV operator basis u13} following the class ordering from Table~\ref{complete CPV operator basis}.

\begin{table}[t]
\centering
\begin{tabular}{|c|c|c|c|c|c|}
\hline 
   \multicolumn{2}{|c|}{$(X^3)$}  & \multicolumn{2}{|c|}{$(\psi^2\phi^3)$}  &  \multicolumn{2}{|c|}{$(\psi^2\phi^2 D)$}  \\
\hline
  $O_{\Tilde{G}GG}$ & $f^{ABC}\widetilde{G}_\mu^{A\nu} G_\nu^{B\rho} G_\rho^{C\mu}$  &  $O_{t\phi}$ & $(\phi^\dagger\phi)(\overline{q}_3 t\Tilde{\phi})$ & $O_{\phi tb}$ & $i( \Tilde{\phi}^\dagger D_\mu \phi ) ( \Bar{t} \gamma^\mu b)$  \\
  $O_{\Tilde{W}WW}$ & $\epsilon^{IJK}\widetilde{W}_\mu^{I\nu} W_\nu^{J\rho} W_\rho^{K\mu}$ & $O_{b\phi}$ & $(\phi^\dagger\phi)(\overline{q}_3 b \phi)$ & & \\
\hline \hline 
  \multicolumn{2}{|c|}{$(X^2\phi^2)$} & \multicolumn{2}{|c|}{$(\psi^4)$} & \multicolumn{2}{|c|}{$(X\psi^2\phi)$} \\
\hline
  $O_{\phi \Tilde{G}}$ & $\phi^\dagger\phi\widetilde{G}_{\mu\nu}^A G^{A\mu\nu}$ & $O_{qtqb}^{(1)}$ & $(\Bar{q}^j_3 t) \epsilon_{jk} (\Bar{q}^k_3 b)$ & $O_{tG}$ & $(\overline{q}_3 \sigma^{\mu\nu}T^A t)\Tilde{\phi}G^A_{\mu\nu}$  \\
  $O_{\phi \Tilde{W}}$ & $\phi^\dagger\phi\widetilde{W}^I_{\mu\nu} W^{I\mu\nu}$ & $O_{qtqb}^{(8)}$ & $(\Bar{q}^j_3 T_A t) \epsilon_{jk} (\Bar{q}^k_3 T_A b)$ & $O_{tW}$ & $(\overline{q}_3 \sigma^{\mu\nu}t)\tau^I\Tilde{\phi}W^I_{\mu\nu}$  \\
  $O_{\phi \Tilde{B}}$ & $\phi^\dagger\phi\widetilde{B}_{\mu\nu} B^{\mu\nu}$  &  &   & $O_{tB}$ & $(\overline{q}_3 \sigma^{\mu\nu}t)\Tilde{\phi}B_{\mu\nu}$ \\
  $O_{\phi \Tilde{W}B}$ & $\phi^\dagger\tau^I\phi\widetilde{W}^I_{\mu\nu} B^{\mu\nu}$  &  &  & $O_{bG}$ & $(\overline{q}_3 \sigma^{\mu\nu}T^A b)\phi G^A_{\mu\nu}$ \\
   & & & & $O_{bW}$ & $(\overline{q}_3 \sigma^{\mu\nu}b)\tau^I \phi W^I_{\mu\nu}$ \\
   & & & & $O_{bB}$ & $(\overline{q}_3 \sigma^{\mu\nu}b) \phi B_{\mu\nu}$ \\
\hline
\end{tabular}
\caption{List of CP-odd dimension-6 operators in our reduced basis under the $U(1)^{13}$ symmetry. }
\label{New CPV operator basis u13}
\end{table}

%% file: signandPOI.tex
We concluded in the previous section that our reduced basis is effective to track leading CP violating effects if we consider $\Lambda^{-2}$ interference effects ($\mathcal{M}_{int,i}$) from one dimension-6 operator at a time. Those interferences are always odd under CP symmetry since we neglect the SM CP phase and therefore do not contribute to CP-even observables , such as the total cross-section of C-even processes like $g g \rightarrow t \Bar{t}$ for example,
\begin{equation}
\begin{split}
    \sigma_{tot}^{g g \rightarrow t \Bar{t}} &= \int d\Phi_{LIPS} \left[ \left| \mathcal{M}_{SM}\right|^2 + 2 \mathcal{R}e \left\{ \mathcal{M}_{SM}^* \mathcal{M}_i \right\} + \mathcal{O}(\Lambda^{-4}) \right] \\
    &= \sigma_{SM}^{g g \rightarrow t \Bar{t}} + 0 + \mathcal{O}(\Lambda^{-4}),
\end{split}
\end{equation}
where $\Phi_{LIPS}$ is the CP-even Lorentz-invariant phase-space. As a result, the operators in Tables~\ref{New CPV operator basis u14}(and \ref{New CPV operator basis u13})  have often been constrained thanks to a part of their $\Lambda^{-4}$ contributions,  $\| \mathcal{M}_{\mathcal{O}_i} \|^2$ in Eq.(\ref{amplitude expansion}). This has been done at LEP \cite{LEP:2003aa} and at LHC \cite{Aaboud:2019nkz,Sirunyan:2017zjc}. Those squared amplitudes are CP-even and do contribute to CP-even observables but are more suppressed in $1/\Lambda$. Therefore, analyzing CP-odd operators with the total cross-section is expected to lead to less stringent constraints on their Wilson coefficients but the main drawback is that they do not test whether CP is actually broken. In general, conventional CP-even observables are not suited to efficiently probe CP violating effects since they present no or small variations from expected SM simulations by relying on $\Lambda^{-4}$-suppressed effects \cite{LEP:2003aa,Aaboud:2019nkz,Sirunyan:2017zjc}.

\begin{figure}[hb!]
    \centering
    \includegraphics[width=0.9\textwidth]{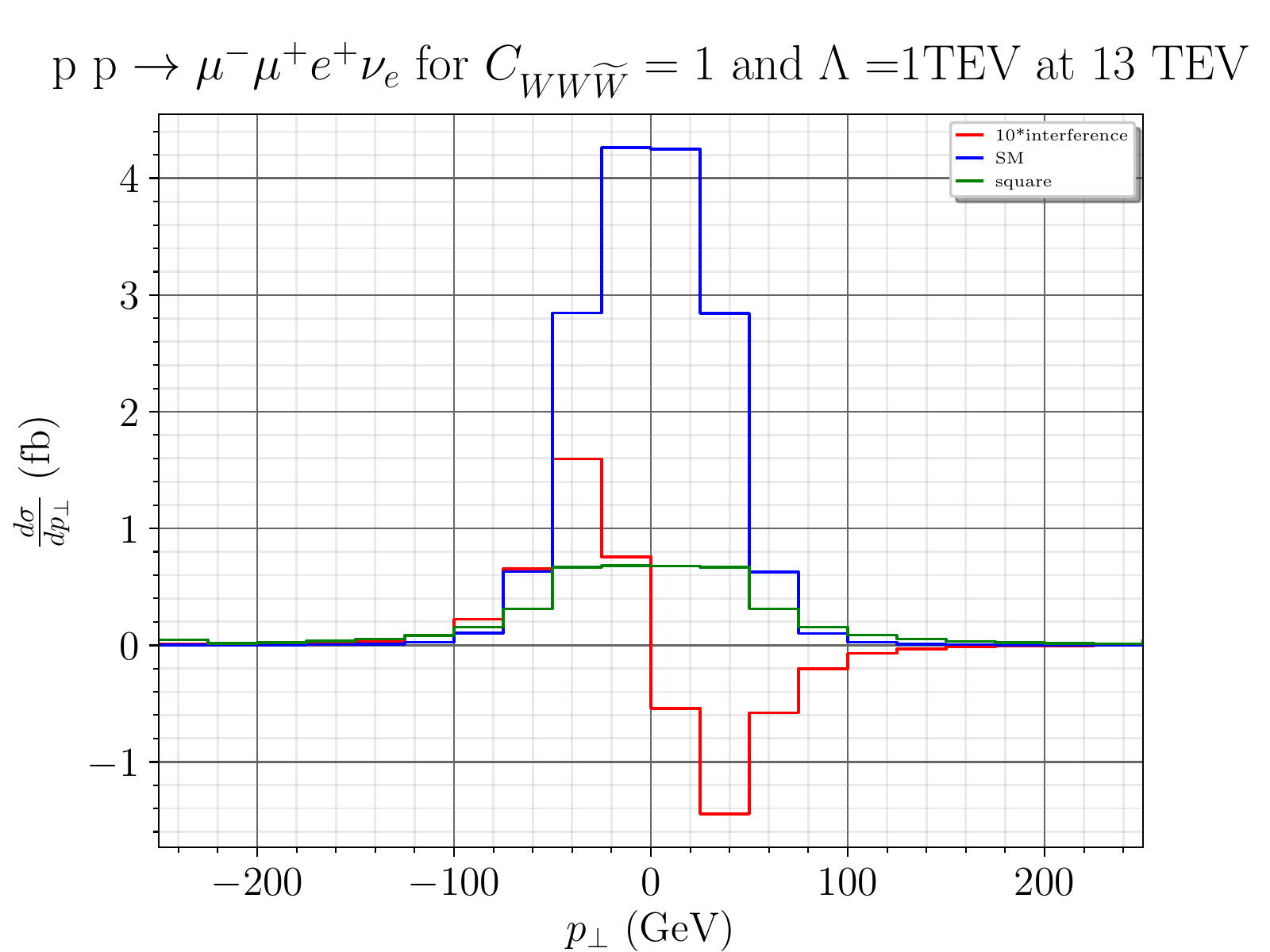}
    \caption{Differential cross-sections of $pp \rightarrow \mu^- \mu^+ e^+ \nu_e$ in ATLAS at $\sqrt{s}=13$ TeV with respect to the triple product $p_\perp$ (the observable is defined and discussed hereafter). The operator considered here is $\mathcal{O}_{\widetilde{W}WW}$ and its Wilson coefficient has been set to 1. The NP scale $\Lambda$ is 1 TeV. %\CD{EpsSign change the sign of the interference}
    }
    \label{fig:tripmom ZW}
\end{figure}

The vanishing interference cross-section is due to flips of the sign of the interference over the phase space. On Figure~\ref{fig:tripmom ZW}, we compare the SM and interference differential cross-sections with respect to an almost CP-odd observable, which will be described in the review below, called the triple product $p_{\perp}$. The interference contribution does not vanish over the whole phase space but actually modulates between positive and negative values, contrary to the SM amplitude which is only positive-definite. The symmetric profile of the pure SM contribution and the asymmetrical profile of initerference are explicit and suggest to use asymmetries to probe CP-odd operators. The relative difference in the heights of the distributions is partially due to the expected $1/\Lambda$ suppression. 
Integrating over the almost CP-odd observable, the positive and negative contributions in the differential cross-section almost completely cancel each other. In C-even process, they would compensate each other exactly and the interference would vanish. On the contrary, when looking at the asymmetry in the CP-odd observable, we get rid of the symmetric CP conserving contributions and highlights asymmetric CP violating effects by keeping track of both positive and negative values of the interference. That is if the asymmetry is defined from a (almost) CP-odd observable. 

We choose to focus on a single asymmetry for each observable/distribution for each process to keep the computation of efficiencies in reviving CP violating effects and their comparisons simple. However, differential asymmetries or distribution could be looked into at a later stage to improve the sensitivities. 

ATLAS is already investigating a CP-odd observable in $Z+jj$ events~\cite{Aad:2020sle}. The simulations of CP-odd operators signals in differential distributions with respect to a CP-odd observable, in this case $\Delta \phi$ presented in the review below, highlight this modulation of the interference amplitude and the phase-space cancellation in CP-even observables for CP-odd operators.

\subsection{Relevant Processes}\label{subsec:POI}

In this section, we summarise the most interesting processes that could directly probe vertices in Table \ref{tab:couplings}. As mentioned above, the $\Lambda^{-2}$ contribution from dimension-six CP-odd operators to the total cross-section of C-even processes vanishes and can only be constrained using differential distributions or, in particular, asymmetries. To measure accurately asymmetries or differential distributions in general, we need a large number of events and therefore processes with quite large cross-section. Hence, we focus on low multiplicity processes.

In the $X^3$ class, the $\mathcal{O}_{\widetilde{G}GG}$ could be investigated by looking at multi-jet events. However, this may not be easy as even the CP-even version of this operator $\mathcal{O}_{GGG}$ is known for its vanishing or suppressed interference \cite{Krauss:2016ely,Hirschi:2018etq,Azatov:2016sqh}. Moreover, this operator is already well constrained by the indirect upper bound measurements of the neutron electromagnetic dipole moment (EDM) (see Ref.~\cite{Dekens:2013zca}). The vertices with the smaller number of legs of $\mathcal{O}_{\widetilde{W}WW}$ are triple gauge boson interactions and contribute to the diboson production at LEP, the Tevatron, and the LHC  (see Refs.\cite{Heister:2001qt, doi:10.1146/annurev.nucl.48.1.33,Falkowski:2016cxu} and Refs.~\cite{Wang:2014uea,Grojean:2018dqj} for comparative studies). This operator suffers the same suppression as the CP-even operator $\mathcal{O}_{WWW}$ due to helicity selection rules, see Ref.~\cite{Azatov:2016sqh}. Since $\mathcal{O}_{\phi\widetilde{W}B}$ has the same vertices, it can be constrained by the same processes. Those two CP-odd operators affect also lower cross-section processes such as weak vector boson fusion (VBF), weak vector boson scattering (VBS) or triple gauge bosons production. Moreover, $\mathcal{O}_{\phi\widetilde{W}}$ and $\mathcal{O}_{\phi\widetilde{B}}$ contribute to the latter with an intermediate Higgs boson.

The operators from the $X^2\phi^2$ class but $\mathcal{O}_{\phi\widetilde{W}B}$ have at least one Higgs in each of their vertices and therefore contribute to Higgs production and decay. However, the $1\to 2$ and $2\to 1$ kinematics does not allowed to build P-odd observables and therefore requires either to add particle (e.g. jets in Higgs production by gluon fusion) or to be sensitive to the polarisation of the vector boson through their decay products or the distribution of the jets in VBF.

The only operator in the $\psi^2\phi^3$ class, $\mathcal{O}_{t\phi}$, is a correction to the SM top quark Yukawa, therefore it definitively affects the top pair production in association with a Higgs. The new measurements from ATLAS and CMS \cite{ATLAS:2020ior,Sirunyan:2020sum} offer new insights into the CP-nature of the Higgs,  rejecting the full CP-odd state, and provide better constraint on $\mathcal{O}_{t\phi}$. This operator also affects at one-loop the production of single Higgs but its interference vanishes by symmetry. The recent observation of $t\Bar{t}t\Bar{t}$ in ATLAS \cite{Aad:2020klt} could also probe this operator.

The last class is the dipole operators $X\psi^2\phi$. These operators generate couplings that can include all type of particles. This means that they appear in a large variety of processes. For example, $\mathcal{O}_{tG}$ contribute to top pair production \cite{Bernreuther:2015yna}, single top production, top pair production in association with a Higgs and Higgs production at one-loop. $\mathcal{O}_{tW}$ is well known for its effects on top decay and single top production while $\mathcal{O}_{tB}$ requires an extra photon or Z boson such as in $pp\to t\bar{t}Z/\gamma$ or $e^+e^-\to t\bar{t}$. 

Nevertheless, they all contribute to EDMs. SM contributions to EDMs arise only beyond the two loop level, hence they are largely suppressed, whereas some of our SMEFT operators contribute already at the one loop level~\cite{Panico:2018hal,Dekens:2013zca}. EDMs offers a big opportunity to constrain CP odd SMEFT operators in indirect observations.

%% file: review.tex
We present here a short review on SMEFT analyses looking for CP violating effects arising from the operators listed in Table \ref{New CPV operator basis u14} and focus on which CP-odd observables are used. Other studies involving CP-odd operators not included in our reduced basis in Table \ref{New CPV operator basis u14} exist. In particular, the rich literature on 4-fermions operators will not be covered here. In the following, we divide measurements into direct and indirect observables. The first tracks CP violating effects arising in high energy processes at colliders while the latter look into deviations in low-energy properties by high energy corrections. 
\\~\\
\underline{Direct observables}:
    \begin{figure}[t]
        \centering
        \includegraphics[scale=0.6]{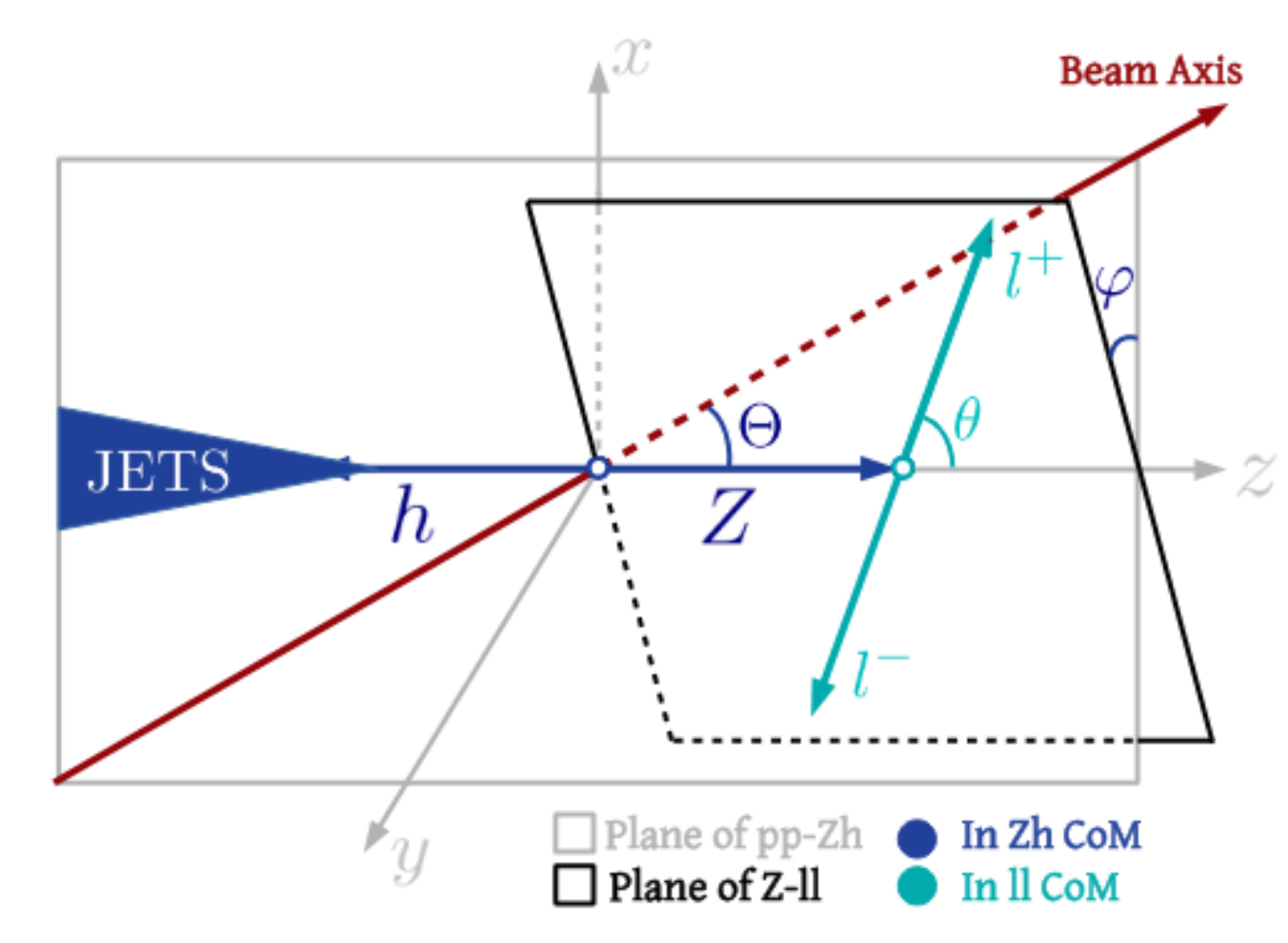}
        \caption{Description of the $Zh$ production in the center-of-mass frame providing guidance on the different frames in which the angles are defined. }
        \label{fig:zh_angles}
    \end{figure} 
\paragraph{Higgs in association with a vector boson:}
In $Zh$ production~\cite{Banerjee:2019pks}, the leading contribution of $\mathcal{O}_{\phi\widetilde{W}B}$, $\mathcal{O}_{\phi\widetilde{W}}$ and $\mathcal{O}_{\phi\widetilde{B}}$ is proportional to  
\begin{equation}\label{eq:zhangle}
    \sin\varphi \sin{2\Theta}\sin{2\theta}
\end{equation}
where the angles are defined in two frames. All angles and frames are displayed in Figure \ref{fig:zh_angles}, taken from Ref.~\cite{Banerjee:2019pks} for better understanding, and defined as follows. The right-handed basis $\{\hat{x}, \hat{y}, \hat{z} \}$ is constructed in the $Zh$ center-of-mass frame such that the $\hat{z}$ axis is aligned with the $Z$ momentum; the $\Hat{y}$ axis is normal to the plane formed by $\hat{z}$ and the beam axis; the $\Hat{x}$ axis completes the set of normal coordinates. $\Theta$ is the angle between the $Z$ momentum and the beam axis and $\varphi$ is the angle between the $Z$ decay products plane and the $(\hat{x}, \hat{z})$ plane, both in the $Zh$ center-of-mass frame. Finally, $\theta$ is the angle of the positively charged lepton momentum with the $\hat{z}$ axis in the $Z$ center-of-mass frame. Since the CP-odd contribution vanishes after integration over the phase space, the constraints are obtained by convolution with the function of the angle in Eq.~\eqref{eq:zhangle}. 
Similarly but for $Wh$ production, the same angle $\varphi$ but defined from the W instead of the Z boson, $\varphi_W$, was used to probe CP violating effects~\cite{Bishara:2020vix}.
\paragraph{Signed azimuthal angle difference:}
The asymmetry in the signed azimuthal angle difference $\Delta \left(\Delta \phi_{pp'} \right)$ between final state particles $p$ and $p'$ is defined as
\begin{equation}
    \Delta \left(\Delta \phi_{pp'} \right) = \sigma(\Delta \phi_{pp'} < \pi/2) - \sigma(\Delta \phi_{pp'} > \pi/2),
\end{equation}
where the final state particles are ordered by their rapidities $y_p$, $y_{p'}$ such that, if $y_{p'}>y_p$, then
\begin{equation} \label{eq: signed azim angle def}
    \Delta \phi_{pp'} = \phi_{p'} - \phi_p.
\end{equation}
and, by construction, $\Delta \phi_{pp'} \in [-\pi,\pi]$. Namely, this is the difference in azimuthal angles between the 'forward' particle and the 'backward' particle. In the context of SMEFT, $\Delta\phi_{jj'}$ was studied in Higgs production by WBF, \textit{i.e.} in the $hjj$ production channel where the signed azimuthal angle difference between the jets is probed, see Refs.~\cite{Plehn:2001nj,Brehmer:2017lrt,Englert:2019xhk}. Ref.~\cite{Englert:2019xhk} further applied the observable in Higgs associated with a top pair production where the signed azimuthal angle difference is between the leptons $l^{ (\prime) }$ originating from the two top decays $\Delta\phi_{ll'}$. Recently, this observable was applied in the leptonic decay of diboson production~\cite{DasBakshi:2020ejz}. For $W^+W^-$, $p$ and $p'$ in $\Delta \phi_{pp'}$ are the two detectable leptons $l^+$ and $l^-$ from the decay of the two massive bosons. In case of a neutral boson, a $Z$ or a photon, associated with a $W$, the momentum of the neutral boson is used with the lepton $l$ from the decay of $W$. In Ref.~\cite{Brehmer:2017lrt} it is mentioned that this observable is related to a triple product but the experimental limitations do not allow to measure the triple product directly with the required precision. 
\paragraph{h $\to$ 4l:}
Another observable has been proven useful to the particular case of a Higgs boson decaying in two same-flavour, opposite-charge lepton pairs through two decaying neutral bosons
\begin{equation}
    p~ p \rightarrow h \rightarrow l^+(p_1)~ l^-(q_1)~ l^{'+}(p_2)~ l^{'-}(q_2),
\end{equation} 
thus testing the $hZZ$ vertex. One can construct the angle $\Phi \in [-\pi, \pi]$ between the $Z$ decay planes in the the single resonance frame
\begin{equation}
    \Phi = \frac{\Vec{P} . (\hat{n}_1 \times \hat{n}_2)}{|\Vec{P} . (\hat{n}_1 \times \hat{n}_2)|} \arccos{(-\hat{n}_1 . \hat{n}_2)}.
\end{equation}
where the vectors $\Hat{n}_i$ are normal to each decay plane, $\Hat{n}_i = \frac{( \Vec{p}_i \times \Vec{q}_i)}{|\Vec{p}_i \times \Vec{q}_i|}$, and $\Vec{P}$ is the 3-momentum of one $Z$ boson.  Rather than the actual value of this angle, it is the sign of $\Phi$ which is the CP sensitive observable or equivalently of $\sin{\Phi}$. Thus, an asymmetry $\Delta \Phi$ or $\Delta \sin{\Phi}$ is again helpful to extract information on CP violating effects. This observable was originally designed to probe the CP and tensor structure of a single-produced resonance decaying in SM particles~\cite{Gao:2010qx,Bolognesi:2012mm} but was later used to constraint SMEFT operators in the four-leptons decay of the Higgs in Refs.~\cite{Beneke:2014sba,Brehmer:2017lrt}, despite the slightly different ways to construct the angle. We emphasize that a triple product is the starting point to derive this observable.
\paragraph{Diboson production:}
A recently developed direct observable has been developed in diboson production 
\begin{equation}
    p~ p \rightarrow W~ Z/\gamma
\end{equation} 
in Refs.~\cite{Azatov:2019xxn,Panico:2017frx}. Particularly Ref.~\cite{Azatov:2019xxn} presents a short development to introduce the observable in $W \gamma$ production that can be translated to $WZ$ production. Considering the $2 \rightarrow 3$ process, the differential cross section is given by 
\begin{equation}
    \frac{d\sigma}{d\Omega} = \frac{1}{2s} \frac{|\sum_{hel=0,\pm} (\mathcal{M}^{SM}_{q\Bar{q}\rightarrow \gamma+ W_{hel}} + \mathcal{M}^{dim6}_{q\Bar{q}\rightarrow \gamma+ W_{hel}}) \mathcal{M}_{W_{hel} \rightarrow l-\Bar{\nu}+}  |^2}{(p_W^2 - m_W^2)^2 + m_Z^2 \Gamma_W^2}
\end{equation}
where $hel$ are the possible helicities for the intermediate W boson and $\Omega=(2\pi)^4 \delta^4( \sum_i p_i - p_f) \Pi_i \frac{d^3 p_i}{2E_i (2\pi)^3}$ is the phase space. $m_Z$ and $m_W$ are the masses of the Z and W bosons respectively, and $\Gamma_W$ is the width of W. In the narrow width approximation, the leading contribution of the interference term reduces to
\begin{equation}\label{eq:interfWA}
    \frac{\pi}{2s} \frac{\delta(s-m_W^2)}{\Gamma_W m_W} \mathcal{M}^{SM}_{q\Bar{q}\rightarrow \gamma+ W_{T_-}} (\mathcal{M}^{dim6}_{q\Bar{q}\rightarrow \gamma+ W_{T_+}})^* \mathcal{M}_{W_{T_-} \rightarrow l-\Bar{\nu}+} (\mathcal{M}_{W_{T_+} \rightarrow l-\Bar{\nu}+})^* + h.c.
\end{equation}
The longitudinal polarizations are subleading contributions so there are not included, neither in Ref.\cite{Azatov:2019xxn} nor here. The interference would vanish due to the helicity selection rules without the last two factors in Eq.(\ref{eq:interfWA}); this is the interference resurrection. If only the $2 \rightarrow 2$ process was considered and the branching ratios were applied, the interference would vanish. The decay amplitudes prevent this from happening: it is only by looking at the complete process that the interference can be appreciated. Moreover, the decay amplitudes are such that
\begin{equation*}
    \mathcal{M}_{W_{T_-} \rightarrow l-\Bar{\nu}+} (\mathcal{M}_{W_{T_+} \rightarrow l-\Bar{\nu}+})^* \propto e^{-2i\phi_W},
\end{equation*}
with $\phi_W$ being the angle between the plane determined by the W decay products and the $W\gamma$ scattering plane. The amplitude produced by the product of the first two terms in Eq.(\ref{eq:interfWA}), 
\begin{equation*}
    P(a\rightarrow b) = \mathcal{M}^{SM}_{q\Bar{q}\rightarrow \gamma+ W_{T_-}} (\mathcal{M}^{dim6}_{q\Bar{q}\rightarrow \gamma+ W_{T_+}})^*
\end{equation*} 
has definite CP-properties
\begin{equation*}
    P(a \rightarrow b) = - P(b \rightarrow a),
\end{equation*}
but simultaneously the optical theorem ensures that the amplitude is equivalent to the conjugate amplitude of the reversed process. Thus,
\begin{equation*}
    P(a \rightarrow b) = -  P(a \rightarrow b)^*.
\end{equation*}
When considering CP-odd dimension-six operators this means that the hermitian conjugate of the interference amplitude will take a minus sign and an opposite phase with respect to the original interference amplitude, resulting in a interference cross section depending on $\sin{\phi_W}$. This development is valid for $WZ$ production as well when applied for each decaying boson. In the end, we have
\begin{equation}\label{eq:barducciobservables}
\begin{split}
    |\mathcal{M}_{W \gamma}|^2 & ~~ \propto \sin{2\phi_W}, \\
    |\mathcal{M}_{W Z}|^2 & ~~ \propto \sin{2\phi_W} + \sin{2\phi_Z}.
\end{split}
\end{equation}
The reader is referred to Ref.~\cite{Panico:2017frx} for a rigorous development. Practically, this observable is constructed similarly to $\Phi$ but, rather than looking at the angle between two vector boson decay planes, one looks at the angle between the decay and scattering planes for each boson. To construct the observable a reference axis is needed. Refs.~\cite{Azatov:2019xxn,Panico:2017frx} use the $\Hat{z}$-axis, the one aligned with the beam direction, in the laboratory frame and we construct the two vectors $\hat{n}_{scat}$ and $\hat{n}_{decay}$ defined as,
\begin{equation}
\begin{split}
     \hat{n}_{scat} &= \frac{\hat{z} \times \Vec{p}_{V}}{|\Vec{p_V}|}, \\
     \hat{n}_{decay} &=\frac{\Vec{p}_{l+} \times \Vec{p}_{l-}}{|\Vec{p}_{l+} \times \Vec{p}_{l-}|},
\end{split}
\end{equation}
where $V=\{W,Z\}$ is the considered vector boson decaying into the leptons $l_+$ and $l_-$, $\pm$ indicates the helicity of the lepton. The two vectors are normal to the scattering plane and the decay plane of the vector boson forming an angle $\phi_V$ between them. The angle is obtained by  
\begin{equation}
    \phi_V = \textrm{sign}[ (\hat{n}_{scat} \times \hat{n}_{decay}) . \Vec{p}_{V}] \arccos{( \hat{n}_{scat} . \hat{n}_{decay} }).
\end{equation}
In the end, the considered observable is the sine of twice the angle between the two planes, $\sin 2\phi_V$. Following the treatment of the amplitudes above and their final dependence in Eq.(\ref{eq:barducciobservables}) the observable is $\sin 2\phi_W$ in $W\gamma$ production or $\sin 2\phi_W + \sin 2\phi_Z$ in $WZ$ production and will probe CP violating effects in diboson production. We note that a triple product is again present to construct this CP-odd observable.
\paragraph{Top pair production:}
The top is expected to be particularly sensitive to NP. Top pair production is therefore especially interesting for SMEFT studies including CPV. We summarise here the results in Ref.~\cite{Bernreuther:2015yna} with a focus on CP-odd observables. 
At the partonic level, the production matrix $R^I$ in top pair production can be decomposed in four terms with respect to the top and anti-top spin spaces 
\begin{equation}\label{eq:partontoppair}
    R^I = f^I [A^I \mathbb{1} \otimes \mathbb{1} + \widetilde{B}^{I+}_i \sigma^i \otimes \mathbb{1} + \widetilde{B}^{I-}_i \mathbb{1} \otimes \sigma^i + \widetilde{C}^I_{ij} \sigma^i \otimes \sigma^j]
\end{equation}
where $\mathbb{1}$ is the $2\times2$ unit matrix and $\sigma^i$ ($\sigma^j$) are the Pauli matrices forming the top (anti-top) spin space. The index $I$ represents the initial state in top pair production which can be either gluon-gluon $(g g)$ or quark-antiquark $(q\Bar{q})$ at LO. An orthogonal basis $\{\Hat{r},\Hat{k}, \Hat{n}\}$ from the top and parton momenta is defined as follows: $\hat{k}$ is aligned with the top quark momentum in the $\bar{t}t$ center-of-mass frame,  $\hat{p}$ is the direction of one of the initial partons in the same frame and $\hat{r}$ and $\hat{n}$ are respectively
\begin{equation*}
    \hat{r} = \frac{1}{r} (\hat{p}-y\hat{k}), ~~ \hat{n}=\frac{1}{r}(\hat{p}\times\hat{k}),
\end{equation*}
where $ y = \hat{p}.\hat{k}$ and $r=\sqrt{1-y^2}$. Namely, $\hat n$ is orthogonal to the scattering plane while  $\hat r$ is in the scattering plane and orthogonal to $\hat k$.
This basis ensures that each energy-dependent component of $\widetilde{B}^{I\pm}_i$ and $\widetilde{C}^I_{ij}$ now respects definite properties under charge conjugation C-, parity P- and pseudo-time $T_N$-transformations. They are decomposed as follows,
\begin{eqnarray*}
    \Hat{B}_{i}^{I\pm} &=& b_{r}^{I\pm} \hat{r}_i + b_{k}^{I\pm} \hat{k}_i + b_{n}^{I\pm} \hat{n}_i , \\
    \Hat{C}_{ij}^{I} &=& c^I_{rr} \hat{r}_i \hat{r}_j + c^I_{kk} \hat{k}_i \hat{k}_j + c^I_{nn} \hat{n}_i \hat{n}_j \\
                    & & + c^I_{rk} (\hat{r}_i \hat{k}_j + \hat{k}_i \hat{r}_j) + c^I_{kn} (\hat{k}_i \hat{n}_j + \hat{n}_i \hat{k}_j) + c^I_{rn} (\hat{r}_i \hat{n}_j + \hat{n}_i \hat{r}_j) \\
                    & & + c^I_r (\hat{k}_i \hat{n}_j - \hat{n}_i \hat{k}_j) + c^I_k (\hat{n}_i \hat{r}_j - \hat{r}_i \hat{n}_j) + c^I_n (\hat{r}_i \hat{k}_j - \hat{k}_i \hat{r}_j).
\end{eqnarray*}
$c_k$, $c_r$ , $b_r^+-b_r^-$ and $b_k^+-b_k^-$ are all CP-odd and P-odd while $c_n$ and $b_n^+-b_n^-$ are CP-odd but P-even and they all can constrain CP-odd operators. These coefficients are negligible in the SM because they arise from the Kobayashi-Maskawa phase or from absorptive phase at one-loop. Since all our operators are P-odd, they can only be constrained by the first four combinations. These coefficients defined at the partonic level are not observables per se and should be linked to actual accessible measurements, \textit{i.e.} independent of the initial partons and dependent on the top decay products: the charged leptons, the jets and the missing transverse energy in semi- or di-leptonic events.
 Fortunately, the charged lepton directions are highly correlated with the top spin direction.
The normalised differential cross section expressed as a function of the angular distributions of the final state leptons, $\Omega_{\pm}=\phi_\pm \cos{\theta_\pm}$, 
\begin{equation}\label{eq:particletoppair}
    \frac{1}{\sigma} \frac{d\sigma}{d\Omega_{+} d\Omega_{-}} = \frac{1}{(4\pi)^2} [1 + \Vec{B}_1 .\Hat{l}_+ + \Vec{B}_2.\Hat{l}_- + \Hat{l}_+.C'.\Hat{l}_-]
\end{equation}
offers therefore an access to the top quarks spins.
In this equation, the vector $\Hat{l}_+$ ($\Hat{l}_-$) is the unit vector aligned with the lepton $l^+$ ($l^-$) direction of flight in the $t$ ($\Bar{t}$) rest frame. 
The definition of two reference axes $\Hat{a}$ and $\Hat{b}$ allows to define the angles
\begin{equation}
    \cos{\theta_+} = \Hat{l}_+ . \Hat{a} ~~\text{and}~ \cos{\theta_-} = \Hat{l}_- . \Hat{b}.
\end{equation}
Integrating the differential cross section over all the other variables, we obtain
\begin{equation}\label{eq:particletoppair}
    \frac{1}{\sigma} \frac{d\sigma}{d\cos{\theta_+} d\cos{\theta_-}} = \frac{1}{4} [1 + B_1(\Hat{a}) \cos{\theta_+} + B_2(\Hat{b}) \cos{\theta_-} + C(\Hat{a},\Hat{b}) \cos{\theta_+} \cos{\theta_-} ].
\end{equation}
The reference axes $\Hat{a}$ and $\Hat{b}$ are chosen from a non-orthogonal set of vectors $\{\Hat{p}_p, \Hat{r}_p, \Hat{n}_p \}$:  $\hat{k}$ is still aligned with the top quark momentum in the $\bar{t}t$ center-of-mass frame, now $\hat{p}_p=(0,0,1)$ is one of the proton beam direction in the laboratory frame, then $\hat{r}_p$ and $\hat{n}_p$ are respectively
\begin{equation*}
    \hat{r}_p = \frac{1}{r_p} (\hat{p}_p-y_p\hat{k}), ~~ \hat{n}_p=\frac{1}{r_p}(\hat{p}_p\times\hat{k}), 
\end{equation*}
where $ y_p = \hat{p}_p.\hat{k}$ and $r_p=\sqrt{1-y_p^2}$. The authors define several sets of reference axes:
\begin{eqnarray*}
    n &:& \hat{a} = sign(y_p)\hat{n}_p, ~ \hat{b} = -sign(y_p)\hat{n}_p, \\
    r &:& \hat{a} = sign(y_p)\hat{r}_p, ~ \hat{b} = -sign(y_p)\hat{r}_p, \\
    k &:& \hat{a} = \hat{k}, ~ \hat{b} = -\hat{k}, \\
    r^* &:& \hat{a} = sign(\Delta|y|)sign(y_p)\hat{r}_p, ~ \hat{b} = -sign(\Delta|y|)sign(y_p)\hat{r}_p, \\
    k^* &:& \hat{a} = sign(\Delta|y|)\hat{k}, ~ \hat{b} = -sign(\Delta|y|)\hat{k},   
\end{eqnarray*}
where $\Delta |y| = |y_t|-|y_{\Bar{t}}|$, $|y|$ being the modulus of the rapidity in the laboratory frame.
The differences $C(n,r) - C(r,n)$ and $C(n,k) - C(k,n)$ (P-odd, CP-odd) probe CP violating effects through correlations between lepton angular distributions. In Table 6 of Ref.~\cite{Bernreuther:2015yna}, we can see that these observables are actually related to the partonic level coefficients $c_k$ and $c_r$, respectively. 
The differences $B_1(k)-B_2(k)$ and $B_1(k^*)-B_2(k^*)$ are related to the difference in transverse polarizations $b^{I+}_{k} - b^{I-}_{k}$ defined at the partonic level. A similar situation appears for $B_1(r)-B_2(r)$  and $B_1(r^*)-B_2(r^*)$ with $b^{I+}_{r} - b^{I-}_{r}$. For these differences the SM contributions vanish at tree-level and any sizeable value would point to CPV from NP.
    
One particularly interesting point is that differences of correlation coefficients $C(n,r) - C(r,n)$ and $C(n,k) - C(k,n)$ are related to CP-odd triple product correlations 
\begin{equation}
\begin{split}
    O^{CP}_1 &=(\Hat{l}_+ \times \Hat{l}_-).\Hat{k}, \\
    O^{CP}_2 &= sign(y_p) (\Hat{l}_+ \times \Hat{l}_-).\Hat{r}_p,
\end{split}
\end{equation}
such that
\begin{equation}
\begin{split}
    C(n,r) - C(r,n) & = 9<O^{CP}_1> , \\
    C(n,k) - C(k,n) & = -9<O^{CP}_2>.
\end{split}
\end{equation}
\paragraph{Triple products:}
   
Genuine triple products are older observables, first used at the Tevatron \cite{Dawson:1996ge}, where quark and anti-quark direction where statistically easier to access due to an anti-proton beam, and in hadron decays (see Ref.\cite{Donoghue1987} for an example in kaon decays). They have been used in diboson production as well. 
    
In $WZ$ production, the triple product $\Vec{p}_q . (\Vec{k}_Z \times \Vec{p}_l)$ where $\Vec{p}_q$ is the 3-momentum of the incoming quark, $\Vec{k}_Z$ the reconstructed 3-momentum of the $Z$ boson and $\Vec{p}_l$ the outgoing charged lepton has been use to probe CP violation due to the anomalous triple coupling $\widetilde{\lambda}_Z$, which is originating from $\mathcal{O}_{\widetilde{W}WW}$, in WZ production~\cite{Kumar:2008ng}. Since the incoming quark momentum is not an observable quantity, the authors replaced it with $(0,0,\Vec{k}_Z^z)$ as a proxy. They find that the replacement holds around 70\% of the time so the sensibility to CP largely remains. Their asymmetry $\Delta$ is then built by weighting the events such that
\begin{equation}
    \Delta = \int d\sigma~ \Xi^z_\pm (k_Z, p_l) .
\end{equation}
where 
\begin{equation}\label{eq: def Xi}
    \Xi^z_\pm (k_Z, p_l) = sign(k_Z^z)~ sign[(\Vec{k}_Z \times \Vec{p}_l)^z].
\end{equation}
    
In $W^+ W^-$ production, Ref.~\cite{Han:2009ra} also considers a genuine triple product $(\Vec{p}_{f} \times \Vec{p}_{\Bar{f}}) . \Vec{p}_q$ with $\Vec{f}$ $\left(\Vec{\Bar{f}}\right)$ being the 3-momentum of the positively (negatively) charged lepton but faces the same issues with the quark momentum $\Vec{p}_q$. Consequently, the triple product is multiplied by a quantity that depends on the quark direction as well, which we will refer as the direction factor. The final observable is quadratic in the beam direction and will not be affected by the uncertainty in the actual quark direction. The observable is noted here as
\begin{equation}
    O_Z = \left( (\Vec{p}_{f} \times \Vec{p}_{\Bar{f}}) . \Vec{p}_q \right) sign[(\Vec{p}_{f} - \Vec{p}_{\Bar{f}}) . \Vec{p}_q]
\end{equation}
and $\Vec{p}_q$ is substituted by the $\Hat{z}$ axis without losing any information on the sign of $O_Z$. This is a generalization of the observable in Ref.~\cite{Dawson:1996ge}.

In $W\gamma$ production, Ref.~\cite{Dawson:2013owa} follows a similar approach and argues that the best observables to probe CPV are the asymmetries in the triple products
\begin{equation}
\begin{split}
    O_W & = (\Vec{p}_\gamma \times \Vec{p}_{beam}) . \Vec{p}_l, \\
    O_{\gamma} &= \Vec{p}_{\gamma} . \Vec{p}_{beam} O_W, \\
    O_{l} &= \Vec{p}_l . \Vec{p}_{beam} O_W.
\end{split}
\end{equation}
Just like in Ref.~\cite{Han:2009ra}, the first observable is ill-defined due initially to the quark direction and returns a zero asymmetry, as opposed to the last two which contain the direction factor and are quadratic in the beam direction. However, due to the $pp$ initial state, the $W^\pm \gamma$ processes are not exactly CP eigenstates from each other and a non-zero asymmetry would not directly provide a proof of CPV. We precise to the reader that Ref.~\cite{Dawson:2013owa} allows for absorptive phases (we do not here). 
    
In $Z\gamma$ production, the asymmetry in $O_Z$ is used to probe CP violating effects in Ref.~\cite{Dawson:2013owa}. Reciprocally to the angles related to triple products above, these triple products determine different angles in the final state. As an explicit example, Ref.~\cite{Han:2009ra} practically defines an azimuthal angle $\Phi$ between the two outgoing charged leptons in the transverse plane,
\begin{equation}
    \Phi = sign[(\Vec{p}_{f} - \Vec{p}_{\Bar{f}}) . \Hat{z}] \arcsin{\left[(\Vec{p}_{f} \times \Vec{p}_{\Bar{f}}) . \Hat{z} \right]} 
\end{equation}
and provides predictions at the LHC for the asymmetry
\begin{equation}
    A_{\Phi} = \frac{ \mathcal{N}_{\Phi>0} - \mathcal{N}_{\Phi<0} }{ \mathcal{N}_{\Phi>0} - \mathcal{N}_{\Phi<0} }.
\end{equation}

\begin{table}
\centering
\begin{tabular}{|c|c|c|c|}
\hline
  Ref  &  Operator(s) &  Observable  & Process \\
\hline
\cite{Panico:2017frx,Azatov:2019xxn} & $\mathcal{O}_{WW\widetilde{W}}$ &  $\sin 2\phi_Z+\sin{2\phi_W}$ & $pp\rightarrow ZW$ \\
                        &    &  $\sin{2\phi_W}$ & $pp\rightarrow W\gamma $   \\
\hline
\cite{Kumar:2008ng} & $\mathcal{O}_{\widetilde{W}WW}$ & $sign[(p_Z)^z] sign[(p_l\times p_Z)^z]$ & $pp\rightarrow ZW$   \\
                    & $\mathcal{O}_{\phi\widetilde{W}}$ &  &\\
                    & $\mathcal{O}_{\phi\widetilde{W}B}$ &  &\\
\hline
\cite{Bernreuther:2013aga,Bernreuther:2015yna} &   $\mathcal{O}_{tG}$     &     $B_{1,2}$ and $O^{CP}_{1,2}$   & $pp \rightarrow t\Bar{t}$   \\
\hline
\cite{Bishara:2020vix} & $\mathcal{O}_{\phi \widetilde{W}}$ & $\sin{\phi_W}$ & $pp\rightarrow Wh $  \\
\hline
\cite{Brehmer:2017lrt} & $\mathcal{O}_{\phi\widetilde{W}}$ & $\Delta \phi_{ll}$ & $pp\rightarrow hqq'$ (WBF)     \\
                       & $\mathcal{O}_{\phi\widetilde{B}}$ & $\Delta \phi_{ll}$ & $pp\rightarrow hZ$  \\
                       & & $\sin\Phi$ & $pp\rightarrow h\rightarrow 4l$   \\
\hline
\cite{Englert:2019xhk} & $\widetilde{O}_g \subset \mathcal{O}_{\phi G}$ & $\Delta \phi_{ll}$ & $pp\rightarrow hqq'$ (WBF)    \\
                       & $i\Bar{t}\gamma_5 t h $ & $\Delta \phi_{jj}$ & $pp\rightarrow tth$   \\
\hline
\cite{Plehn:2001nj} & $\mathcal{O}_{\phi\widetilde{W}}$ & $\Delta \phi_{jj}$  & $pp\rightarrow hqq'$ (WBF)   \\
\hline
\cite{Beneke:2014sba}    & $\mathcal{O}_{\phi \widetilde{W}}$  & $\sin{\Phi}$ & $pp \rightarrow h \rightarrow 4l$  \\
                          & $\mathcal{O}_{\phi \widetilde{B}}$ & $\sin{2\phi}$& \\
                          & $\mathcal{O}_{\phi \widetilde{W}B}$ &  &\\
\hline
\cite{Bernlochner:2018opw} & $\mathcal{O}_{\phi \widetilde{G}}$  & $\Delta \phi_{ll}$ & $pp \rightarrow h \rightarrow ZZ^* / \gamma\gamma$  \\
                          & $\mathcal{O}_{\phi \widetilde{W}}$  &  
                          %sensitivity enhanced with $\cos{\Phi}$ 
                          &\\
                          & $\mathcal{O}_{\phi \widetilde{B}}$ & & \\
                          & $\mathcal{O}_{\phi \widetilde{W}B}$ & & \\
\hline
\cite{Banerjee:2019pks} &  $\mathcal{O}_{\phi\widetilde{B}}$  &  $\sin{\varphi}$& $pp \rightarrow Z h$    \\
                        &  $\mathcal{O}_{\phi\widetilde{W}}$  &   & \\
                        &  $\mathcal{O}_{\phi\widetilde{W}B}$ &  &  \\
\hline
\cite{Biekotter:2020flu} &  $\mathcal{O}_{\phi\widetilde{W}}$    &  \multirow{2}{*}{ $\frac{\Vec{p}_\gamma . (\Vec{p}_{j2} \times \Vec{p}_{bb})}{|\Vec{p}_\gamma| |\Vec{p}_{j2}| |\Vec{p}_{bb}|} $  } &   $pp \rightarrow h(\rightarrow b\Bar{b})\gamma jj$ (WBF)    \\
                    & $\mathcal{O}_{\phi\widetilde{W}B}$ &  &  \\
\hline
\cite{DasBakshi:2020ejz}  &   $\mathcal{O}_{\phi \widetilde{W}B}$    &    $\Delta \phi_{Zl}$      &    $p p \rightarrow WZ$ \\
                        & $\mathcal{O}_{\widetilde{W}WW}$    &    $\Delta \phi_{ll'}$      & $pp\rightarrow WW$ \\
                       &      &    $\Delta \phi_{\gamma l}$      &    $pp \rightarrow W\gamma$ \\
\hline
\end{tabular}
\caption{Summary table of direct observables to constrain operators from reduced basis. }
\label{tab:directobs}
\end{table}

The different analyses and their respective observables are gathered in Table \ref{tab:directobs}. In the second column we list which operators are relevant in the respective analyses. For Ref.~\cite{Bernreuther:2015yna}, we only mention the dipole operator $\mathcal{O}_{tG}$ in Table \ref{tab:directobs} even though the authors also considered the two operators
\begin{equation}
\begin{split}
    \mathcal{O}_{gt} &= [\Bar{t}_R \gamma^\mu T^A D^\nu t_R] G^A_{\mu\nu},  \\
    \mathcal{O}_{gQ} &= [\Bar{Q}_L \gamma^\mu T^A D^\nu Q_L] G^A_{\mu\nu}.
\end{split}
\label{eq:aachenop}
\end{equation}
Those two operators can be re-written following the development in Eq.(6.6) in Ref~\cite{Grzadkowski:2010es} in term of the chromomagnetic operator\footnote{and four-fermion operators for their CP-even part.}. This development is independent whether it is done before or after EW symmetry breaking. In the latter case, the full (\textit{i.e.} containing the Higgs field) chromomagnetic operator is replaced by its dimension five version and is proportional to the top mass. We checked analytically that all the top observables introduced in Ref.~\cite{Bernreuther:2015yna} only receive contributions proportional to those of the chromomagnetic operator from the two operators in Eq.~\eqref{eq:aachenop}. In particular, there is no contribution to the CP-odd but P-even observable $b_n^+-b_n^-$ in the SMEFT. Therefore, we confirm that only the chromomagnetic operators belongs to the SMEFT basis and not the two other operators in agreement with Ref.~\cite{Grzadkowski:2010es}.

An interesting point to note is, in most of the CP sensitive observables we have listed, a triple product is either present in the construction of the observable or can be associated with the observable. As expected, triple products asymmetries are a key ingredient to track CP violating effects from NP in many LHC processes.

~~\\
\underline{Indirect observables}:

Indirect measurements are based on low-energy observations and arise from the running of Wilson coefficients from high energy scales to lower energy scales. The SMEFT is used above the EW scale but then the heavy SM particles are integrated out as we go down to electron mass. Many low-energy observables have the capacity to test the CP symmetry. Although, a lot of them investigate operators not present in the reduced basis. For instance, tree-level contributions to kaon decays involve 1st and 2nd generation fermions discarded within our massless fermions hypothesis, or look at contact interactions between the fermions thus investigating 4-fermions operators absent of the reduced basis as well. In the end, among low-energy observables, only EDMs allows to strongly constrain the operators of the reduced basis so far. 

We focus here on the results of Ref~\cite{Panico:2018hal} based on the electron EDM. It should be noted that with our $U(1)^{14}$ symmetry, no EDM can be generated. However, if this symmetry makes sense for the LHC, it is not justified for this observable. Therefore operators outside our reduced basis are also constrained by this measurement.

Below the EW scale, the electron EDM is induced by 
\begin{equation}
    \mathcal{O}_{e \gamma} = \Bar{e}_L \sigma^{\mu\nu} e_R F_{\mu\nu},
\end{equation}
while above this scale, the dipole operator origins from two operators of the Warsaw basis , $\mathcal{O}_{e B}$ and $\mathcal{O}_{e W}$ defined in Table \ref{complete CPV operator basis}, at the tree-level
\begin{equation}
    C_{e\gamma} (m_W) = \frac{v}{\sqrt{2} \Lambda} \left( \sin{\theta_W} C_{eW}(m_W) - \cos{\theta_W} C_{eB}(m_W) \right) + O(\Lambda^{-3}),
\end{equation}
such that the electron EDM is given by 
    \begin{equation}
        d_e(\mu)= \frac{\sqrt{2} v}{\Lambda^2} \text{Im}[\sin{\theta_W} ~ C_{eW}(\mu) - \cos{\theta_W} ~ C_{eB}(\mu)],
    \end{equation}
where $\theta_W$ is the weak angle. Only the CP-odd operators proportional to the imaginary parts of the two operators coefficients contribute as expected. On the contrary, the real parts contribute to the magnetic dipole. Thanks to helicity selection rules, only a few operators mix with those operators at 1-loop and therefore only three operators of our reduced basis are relevant for the 1-loop renormalisation group equations: 
\begin{equation}\label{eq: running op}
    \frac{d}{d \ln{\mu}} \text{Im}
    \begin{pmatrix}
        C_{eW}\\
        C_{eB}
    \end{pmatrix}
    = -\frac{y_e g}{16 \pi^2}
    \begin{pmatrix}
        0 & 2 \tan{\theta_W}(Y_l + Y_e) & \frac{3}{2}  \\
        1 & 0 & \tan{\theta_W}(Y_l + Y_e)
    \end{pmatrix}
    \begin{pmatrix}
        C_{\phi\widetilde{W}} \\
        C_{\phi\widetilde{B}}  \\
        C_{\phi\widetilde{W}B}
    \end{pmatrix}.
\end{equation}
In Eq.(\ref{eq: running op}), $y_e$ is the electron Yukawa and $Y_l$ and $Y_e$ are the hypercharges of left-handed lepton $l$ and right-handed electron $e$. In addition, $\mathcal{O}_{\widetilde{W}WW}$ gives a finite contribution, one that does not evolve with the scale, which is  
\begin{equation}
    \text{Im} C_{eW} = \frac{3}{64\pi^2} y_e g^2 C_{\widetilde{W}WW}.
\end{equation}

\begin{table}[t]
    \centering
    \begin{tabular}{c|c|c}
        Operator & $\Lambda=10$TeV & $\Lambda=1$TeV \\
    \hline
       $\mathcal{O}_{\widetilde{W}WW}$  & $6.4~10^{-2}g^3$ & $1.7~10^{-4}$ \\
       $\mathcal{O}_{\phi\widetilde{W}}$  & $4.7~10^{-3}g^2$ &  $2.0~10^{-5}$\\
       $\mathcal{O}_{\phi\widetilde{B}}$  & $5.2~10^{-3}g^{'2}$ & $6.7~10^{-6}$ \\
       $\mathcal{O}_{\phi\widetilde{W}B}$  & $2.4~10^{-3}g'g$ & $5.6~10^{-6}$ \\
       $\mathcal{O}_{tW}$  & $6.9~10^{-3}y_t g$ & $4.2~10^{-5}$ \\
       $\mathcal{O}_{tB}$  & $1.2~10^{-2}y_t g'$ & $4.0~10^{-5}$ 
    \end{tabular}
    \caption{Constraints on the Wilson coefficients corresponding to the operators listed in the first column from Ref.\cite{Panico:2018hal}. $g$ and $g'$ are respectively the $SU_L(2)$ and $U_Y(1)$ coupling constants, $y_t$ is the top Yukawa coupling constant. The second column basically reproduces the results detailed in Ref.\cite{Panico:2018hal} thus considering $\Lambda=10$TeV. In the last column, we translate the constraints for $\Lambda=1$TeV with the numerical values of the couplings  inserted.%, which are taken at the 2-loop level in the $\overline{MS}$-renormalization scheme from Ref.\cite{Buttazzo:2013uya}. 
    }
    \label{tab:EDMconstraints}
\end{table}

At the 2-loop level, one must take into account 1-loop mixing of the operators already present at the 1-loop level and direct 2-loop contributions to $\mathcal{O}_{eW}$ and $\mathcal{O}_{eB}$. In particular, the electroweak dipole operators of our reduced basis contribute to the electric dipole at this level. Finally, $\mathcal{O}_{\widetilde{W}WW}$ also mixes with $\mathcal{O}_{\phi\widetilde{W}}$, $\mathcal{O}_{\phi\widetilde{B}}$ and $\mathcal{O}_{\phi\widetilde{W}B}$. 

Those result in the strongest constraints on the operators that we consider later in our analysis. They are displayed in Table \ref{tab:EDMconstraints}.

\begin{table}[!h]
\centering
\begin{tabular}{|c|c|c|c|}
\hline
  Ref  &  Op(s) &  Level & Observable   \\
\hline
\cite{Dekens:2013zca} & $\mathcal{O}_{GG\widetilde{G}}$ & 1-loop & $d_n \leq 2.9~10^{-13}~e.fm$  \\
                      & $\mathcal{O}_{WW\widetilde{W}}$ & 1-loop  &          \\
                      & $\mathcal{O}_{\phi \widetilde{G}}$ & 1-loop &     \\
                      & $\mathcal{O}_{\phi \widetilde{W}}$ & 1-loop  &      \\
                      & $\mathcal{O}_{\phi \widetilde{B}}$ & 1-loop  &     \\
                      & $\mathcal{O}_{\phi \widetilde{W}B}$ &  1-loop  &     \\
\hline
\cite{Panico:2018hal} & $\mathcal{O}_{WW\widetilde{W}}$ & 2-loop  & $d_e<1.1~10^{-29}~e.cm$    \\
                      &                                 & + 1-loop finite &       \\     
                      & $\mathcal{O}_{\phi \widetilde{W}}$ & 1-loop   &                      \\
                      & $\mathcal{O}_{\phi \widetilde{B}}$ &  1-loop &      \\
                      & $\mathcal{O}_{\phi \widetilde{W}B}$ & 1-loop  &       \\
                      & $\mathcal{O}_{u W}$ & 2-loop   &                \\
                      & $\mathcal{O}_{u B}$ & 2-loop   &                   \\
\hline
\end{tabular}
\caption{List of the operators of our reduced basis with the type of their contribution to the EDM, including the references of their computations.}
\label{tab:indirectobs}
\end{table}

~~\\
\underline{Global analyses}:

Recent studies combine direct and indirect observables resulting in consistent stringent constraints on Wilson coefficients of CP-odd operators. However, the best constraints on CP-odd operators are actually obtained from EDMs \cite{Fuchs:2020uoc, Panico:2018hal, Cirigliano:2016njn, Cirigliano:2016nyn}. The LHC is not competitive compared to EDMs because of the accidental suppression of SM contributions in EDMs, the precision of its experimental measurements and the good control of the theoretical uncertainties. Nevertheless, direct CP-odd observables still provide highly valuable information to probe blind directions even if large cancellations are necessary to satisfy the EDM constraints. Therefore, rather than being in a competition, direct and indirect observables are complementary approaches to detect new CP violating effects from any CP-odd operator. %These efforts push towards a complete analysis of the set of CP-odd operators in Table \ref{complete CPV operator basis} and CP-even operators which is the long term ultimate objective of SMEFT analyses. 

%% file: analysis.tex
We focus here on the case of diboson production with a $W$ boson and a neutral boson, $Z$ or $\gamma$. We choose to decay each massive gauge boson into leptonic channels with different flavours: $W\rightarrow e^\pm \nu$ and $Z\rightarrow \mu^-\mu^+$. There are four reasons for this choice. Firstly, those channels, even if they are not C-even processes, almost behave as such. Namely, the interference cross-section is heavily suppressed due to a cancellation between the different regions of the phase space. Therefore our aim is to test which observables could disentangle those regions and estimate their efficiency in doing so by using the sign of the matrix element as proposed in Ref.\cite{Degrande:2020tno}. We will propose our own observables but also compare them to those of previous studies. Secondly, those processes have been measured at different center-of-mass energies at the LHC \cite{Aad:2012twa, Aad:2016ett, Aaboud:2016yus, Khachatryan:2016tgp, Khachatryan:2016poo, Aaboud:2019gxl, Sirunyan:2019bez} and the cross-sections are relatively large resulting in expected large numbers of events which are necessary to measure accurately CP-odd observables such as asymmetries. Thirdly, these final states can be easily reconstructed resulting in a quite clean signal and a relatively low background. Finally, their leptonic channels contain only one neutrino compared to the C-even process $W^+ W^-$ and the two different lepton flavours ensure that there is no confusion between the Z and W decays product\footnote{We ignore detector effects such as misidentification.} which makes our analysis easier. We leave the study of other leptonic decays for future work as well as the semi-leptonic and hadronic decays. 

%Feynman Diagrams
\begin{figure*}[t]
    \centering
    \begin{subfigure}[t]{0.3\textwidth}
    \begin{tikzpicture}
    \begin{feynman}
        \vertex (f1) {\(q\)};
        \vertex [below right=1.75cm of f1] (a);
        %\vertex [below left=of a] (f2) {\(q'\)};
        \vertex [below =2.5cm of f1] (f2) {\(q'\)};
        \vertex [right=of a] (c);
        \vertex [right=4.1cm of f1] (b1) {\(Z/\gamma\)};
        \vertex [right=4.1cm of f2] (b2) {\(W\)};
 
        \diagram* {
        (f1) -- [fermion] (a) ,
        (a) -- [fermion] (f2) ,
        (a) -- [boson, edge label=\(W\)] (c),
        (c) -- [boson] (b1),
        (c) -- [boson] (b2),
        };
    \end{feynman}
    \end{tikzpicture}
    \caption{
    %SM s-channel of $WZ/\gamma$ production.
    }
    \end{subfigure}
    ~~~
    \begin{subfigure}[t]{0.3\textwidth}
    \begin{tikzpicture}
    \begin{feynman}
        \vertex (f1) {\(q\)};
        \vertex [right=of f1] (a);
        \vertex [right=of a] (b1) {\(Z/\gamma\)};
        \vertex [below=2.3cm of a] (c);
        \vertex [below=2.3cm  of f1] (f2) {\(q'\)};
        \vertex [right= of c] (b2) {\(W\)};
        \diagram* {
        (f1) -- [fermion] (a) ,
        (a) -- [boson] (b1) ,
        (a) -- [fermion] (c) ,
        (c) -- [fermion] (f2) ,
        (c) -- [boson] (b2)
        };
    \end{feynman}
    \end{tikzpicture}
    \caption{
    %SM t-channel of $WZ/\gamma$ production.
    }
    \end{subfigure}
    ~
    \begin{subfigure}[t]{0.3\textwidth}
    \begin{tikzpicture}
    \begin{feynman}
        \vertex (f1) {\(q\)};
        \vertex [below right=1.75cm of f1] (a);
        \vertex [below =2.5cm of f1] (f2) {\(q'\)};
        \vertex [blob,right=1.15cm of a] (c) { };
        \vertex [right=4.1cm of f1] (b1) {\(Z/\gamma\)};
        \vertex [right=4.1cm of f2] (b2) {\(W\)};
        \diagram* {
        (f1) -- [fermion] (a) ,
        (a) -- [fermion] (f2) ,
        (a) -- [boson, edge label=\(W\)] (c),
        (c) -- [boson] (b1),
        (c) -- [boson] (b2),
        };
    \end{feynman}
    \end{tikzpicture}
    \caption{
    %NP s-channel of $WZ/\gamma$ production.
    }
    \end{subfigure}
    \caption{Feynman diagrams for diboson production in the SM (the s-channel in (a) and the t-channel in (b), u-channel is not displayed) and with the new $WWZ/\gamma$ vertex from the dimension-six operators $\mathcal{O}_{\widetilde{W}WW}$ and $\mathcal{O}_{\phi \widetilde{W}B}$ in (c). }
    \label{ZW production}
\end{figure*}
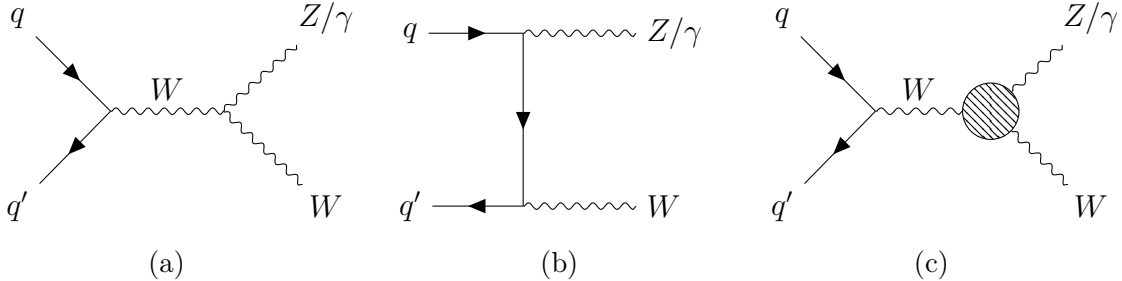

Out of the list of 10 operators from Table~\ref{New CPV operator basis u14}, the relevant operators in $WZ$ and $W\gamma$ production are $\mathcal{O}_{\widetilde{W}WW}$ and $\mathcal{O}_{\phi \widetilde{W}B}$. Those operators modify the WWZ/$\gamma$ coupling present in the s-channel diagram. As a matter of fact, no CP-odd dimension-six contribution arises in the light quark interaction or in the weak bosons decay thanks to our simplification in Section~\ref{subsec:basis reduction u14}. The Feynman diagrams are represented in Figure~\ref{ZW production}. 
The interference between the SM and $\mathcal{O}_{\widetilde{W}WW}$ to the total cross-section of the 2$~\rightarrow~$2 process is suppressed by the helicity selection rules at LO in the massless limit as demonstrated in Ref.~\cite{Azatov:2016sqh}. The same method applied to $\mathcal{O}_{\phi\widetilde{W}B}$ shows that helicity selection rules also suppress its interference\footnote{ Looking at Table II in Ref.~\cite{Azatov:2016sqh} and applying Eq.(9) with the given values results in $h\left(A^{\mathcal{O}_{\phi\widetilde{W}B}}_4\right)=2$ while $h\left(A^{SM}_4\right)=0$.  }. However, at least one of the massive bosons has to be off-shell and, when considering the complete 2$~\rightarrow~$3 or 2$~\rightarrow~$4 processes, the interference is recovered. This is the interference resurrection carefully demonstrated in Ref.~\cite{Azatov:2019xxn,Panico:2017frx} with $\mathcal{O}_{\Tilde{W}WW}$ and reproduced in Eq.(\ref{eq:interfWA}) for W$\gamma$.

We limit our analysis at the leading order (LO) and leave next-to-leading (NLO) corrections for further work. We also limit ourselves to look at the processes at the parton level. It will be insightful to observe the modification in the efficiencies of the observables and their asymmetries due to shower and detector effects but we keep these important aspects for future analyses. This simple framework allows to test many new observables as it is faster, computationally cheaper and easier to understand the results.

The set of operators in Table~\ref{New CPV operator basis u14} has been implemented in a FeynRules model \cite{Alloul:2013bka} where, in both $X^2\phi^2$ and $\phi^3\psi^2$ classes definition, we have subtracted $v^2$ from $(\phi^\dagger\phi)$ for convenience. A UFO model was generated \cite{Degrande:2011ua} and was passed to MG5\_aMC@NLO \cite{Alwall:2014hca}. The PDF set exploited in the event generation is NNPDF2.3, presented in Ref.\cite{Ball:2012cx}, in which $\alpha_S(M_Z)=0.119$. We fix the SM parameters at the Z pole mass :
\begin{equation*}
\begin{split}
    & m_Z=91.1876~ \text{GeV}, ~~  (\alpha_{EM})^{-1}=127.9,~~ G_F= 1.166370~10^{-5}~ \text{GeV}^{-2},\\
    & \Gamma_Z = 2.4952~ \text{GeV},~~ \Gamma_W = 2.085~ \text{GeV}. 
\end{split}
\end{equation*}
The CKM matrix $V_{CKM}$ is reconstructed using the Euler angles, respectively
\begin{equation*}
    \theta_{12} = 0.227~rad, ~~~ \theta_{23}=0.041~rad, ~~~ \theta_{13} = 0.003~rad,
\end{equation*} 
and the CP angle $\delta_{13}=1.2~rad$ in the standard parameterisation of $V_{CKM}$,
\begin{eqnarray*}
 V_{CKM} &=& 
 \begin{pmatrix}
 c_{12} c_{13} & s_{12} c_{13} & s_{13} e^{i\delta_{13}} \\
 -s_{12} c_{23} - c_{12} s_{23} s_{13} e^{i\delta_{13}} & -c_{12} c_{23} - s_{12} s_{23} s_{13} e^{i\delta_{13}} & s_{23}c_{13} \\
 s_{12} s_{23} - c_{12} c_{23} s_{13} e^{i\delta_{13}} & c_{12} s_{23} - s_{12} c_{23} s_{13} e^{i\delta_{13}} & c_{23} c_{13}
 \end{pmatrix}
\end{eqnarray*}
where $c_{ij} = \cos{\theta_{ij}}$ and $s_{ij} = \sin{\theta_{ij}}$. While the CP violating phase is included, we checked that it does not change significantly our results. 

In this work, we focus on the dileptonic decay $W^\pm Z \rightarrow \mu^+ \mu^- e^\pm \overset{\textbf{\fontsize{2pt}{2pt}\selectfont(---)}}{\nu}_{e}$ and the leptonic decay $W^\pm \gamma \rightarrow \gamma e^\pm \overset{\textbf{\fontsize{2pt}{2pt}\selectfont(---)}}{\nu}_{e}$.
We produce 800.000 events for the different final states and each contribution, \textit{i.e.} for the SM, the interference and the squared amplitude to obtain accurate differential distributions. We apply the cuts of the fiducial phase space corresponding to the ATLAS analysis at 13 TeV ~\cite{Aaboud:2016yus} for $WZ$ events:
\begin{equation*}
\begin{split}
    & p_T(\mu)> 15\text{GeV},~~~ |\eta(\mu)|<2.5,~~~  p_T(e)> 20\text{GeV},~~~  |\eta(e)|<2.5, \\
    & \Delta R(\mu^+\mu^-)>0.2,~~~  \Delta R(e\mu^-)>0.3,~~~  \Delta R(e\mu^+)>0.3,\\
    & |m_{\mu^-\mu^+}-m_Z|<10\text{GeV},~~~  m_T(e\nu_e)>30\text{GeV}. 
\end{split}
\end{equation*}

For $W\gamma$ events, we replicate the fiducial phase space corresponding to the ATLAS analysis~\cite{Aad:2012mr} as well. The cuts are :
\begin{equation*}
\begin{split}
    & p_T(e)>25\text{GeV},~~~  |\eta(e)|<2.47,~~~  E_T(\gamma)>15\text{GeV},~~~  |\eta(\gamma)|<2.37, \\
    & E^{miss}_T>35\text{GeV},~~~  \Delta R(e\gamma)>0.7,~~~  m_T(e\nu_e)>40\text{GeV}, \\
    & |m(e\gamma) - m_Z|>10\text{GeV}. 
\end{split}
\end{equation*} 
The cuts for the muon decay of the W boson are very similar except that the last cut is not applied.

\subsection{Formalism}
Searches for the $O_{WW\widetilde{W}}$ have been optimised for the high energy tail of the distributions, as, for example, in Ref.~\cite{Azatov:2019xxn}, to take advantage of the energy growth. Here, we decide to follow a complementary approach by looking at asymmetries over the whole fiducial space. Accurate asymmetry measurements require a high number of events which is not available in high energy tails and therefore we will use instead the large amount of events available close to threshold. 

Consistency with the theoretical development provided in Section~\ref{sec:theory} requires that the differential cross-section with respect to a CP-odd observable $X$ is truncated at the $\Lambda^{-2}$ order such that
\begin{equation}
    \frac{d\sigma}{dX} = \frac{d\sigma(SM)}{dX} +  \frac{c_i}{\Lambda^2}\frac{d\sigma(\obothi)}{dX}.
\end{equation}
Sometimes, we will use the square of the amplitude containing the dimension-six operator but as a mean to check the perturbative behaviour in our analysis and to understand the phase-space suppression of the interference cross-section compared to the SM and the dimension-six cross-sections. %\JT{Mention Schwartz Bound ?}later

We define the asymmetry in $X$ as
\begin{equation}\label{eq:genericdiff}
    \Delta X \equiv  \sigma(X > 0) - \sigma(X < 0) = \Delta X(SM)+  \frac{c_i}{\Lambda^2} \Delta X (\obothi),
\end{equation}
with $\sigma(X>0)=\int_{0}^{b_+} \frac{d\sigma}{dX}~dX$ and $\sigma(X<0)=\int^{0}_{b_-} \frac{d\sigma}{dX}~dX$. $b_\pm$ are the upper and lower bounds of integration of the variable $X$. One could normalise it by taking the ratio of the asymmetry with the total cross-section, 
\begin{equation}\label{eq:genericasym}
    A_{X} \equiv \frac{ \sigma(X>0) - \sigma(X<0) }{  \sigma(X>0) + \sigma(X<0) } = \frac{\Delta X}{\sigma(SM)+ \frac{c_i}{\Lambda^2} \sigma(\obothi)}.
\end{equation}
However, the denominator in the normalised asymmetry is the full cross-section which corresponds to sum of the SM cross-section $\sigma(SM)$ and the contribution of any operator CP violating or not, $\sigma(\obothi)$ and therefore introduce many other parameters dependence. Moreover, the denominator is already quite constrained by the measurements of the total cross-section which are in agreement with the SM prediction. Although we will present our results mainly in term of the asymmetry to avoid those extra dependencies, we will use the SM cross-section value to provide a sensitivity estimate later on.  

We follow the method defined in Ref.\cite{Degrande:2020tno} to quantify the interference suppression over the phase spaces and the efficiencies of our asymmetries.
Namely, we compute the integral of the absolute value of the interference,
\begin{equation}
    \sigma^{|int|} \equiv \int d\Phi \left| \frac{d\sigma_{int}}{d\Phi }\right| 
\end{equation}
from the sum of the absolute value of the weights to understand how much the suppression of the interference is due to the sign flips over the phase space.
The best asymmetry experimentally measurable is given by the measurable absolute value cross-section,
\begin{equation}
    \sigma^{|meas|} \equiv \int d\Phi_{meas} \left|\sum_{\{um\}}\frac{d\sigma_{int}}{d\Phi } \right|
\end{equation}
where the sum runs over the set of unmeasurable quantities $\{um\}$:
\begin{itemize}
    \item the side where the quark and antiquark are coming from;
    \item the flavours of the quarks;
    \item the polarizations of the different final leptons, photons and initial quarks;
    \item the longitudinal momentum of the neutrino.
\end{itemize} 
In this last case, the sum is replaced by the integral over the longitudinal component of the neutrino momentum. This quantity, $\sigma^{|meas|}$, is obtained by integrating and summing the interference matrix element over the unmeasurable quantities for each event generated according to this same interference. Although this is the best measurable asymmetry, it would be computationally expensive to compute especially on real events where transfer functions have to be taken into account. Therefore, we compare additional asymmetries built from simple observables to this one to quantifies their efficiency in order to find simple and efficient observables to constrain the CP violating operators.

\subsection{Triple products}

Our asymmetries are built from triple products in order to improve the sensitivity to CP-odd operators. The triple products require to include the initial quark momenta as well as the momenta of the photon and one of the decay product of the W boson to be non-vanishing in the case of W$\gamma$ production. However, as we will see, including the initial quark and the neutral gauge boson also leads to larger asymmetries in WZ production. Therefore, our triple product is defined as 
\begin{equation}\label{eq:tripmomdef}
    p_\perp (p_l, \Hat{n}_{ref}) = \Hat{n}_{ref} . (\Vec{p}_l \times \Vec{p}_V)
\end{equation}
where $\Vec{p}_l$ is the 3-momentum of one visible lepton, i.e. a electron or a muon, $\Vec{p}_V$ is the reconstructed 3-momentum of the Z boson in $WZ$ production or of the photon in $W\gamma$ production and $\Hat{n}_{ref}$ will be a reference axis approximating the quark momenta. 
In $WZ$ production, the asymmetry $\Delta {p_\perp}(p_{\mu^+}, \Hat{n}_{ref})$ will not be presented in the following because, by definition, it is always opposite to $\Delta {p_\perp}(p_{\mu^-}, \Hat{n}_{ref})$.

The direction of the quark is not experimentally available. However, due the to PDF, quarks are on average more energetic than anti-quarks. To take advantage of this, we explore multiple approximations of the quark momentum $\Hat{n}_{ref}$: the laboratory axis $\Hat{z}=(0,0,1)$, the longitudinal momentum of neutral boson  $\vec n_{Z/\gamma}=(0,0,p^z_{Z/\gamma})$, of the electron $\vec n_e=(0,0,p^z_{e})$ and of the sum of visible particles $\vec n_\sum=(0,0,p^z_{\sum})$ where $p^z_{\sum}$ is the third spatial component of the sum in the momentum of the visible particles. All vectors are taken in the laboratory frame.

Note that some observables mentioned in Subsection~\ref{subsec:review} are equivalent to some of triple product configurations defined here. The triple product defined in Ref.~\cite{Kumar:2008ng}, listed in the review in Eq.\eqref{eq: def Xi}, for $WZ$ production corresponds to 
\begin{equation} \label{eq:Xi}
     \Xi = p_\perp (p_l, \vec n_z).
\end{equation}
In $W\gamma$ production, the observables in Ref.~\cite{Dawson:2013owa} have the same sign as some triple products and will produce equal asymmetries :
\begin{equation*}
\begin{split}
    \mathcal{O}_W & \sim p_\perp (p_e, \Hat{z}), \\
    \mathcal{O}_\gamma & \sim p_\perp(p_e,p_\gamma), \\
    \mathcal{O}_l & \sim p_\perp(p_e,p_e).
\end{split}
\end{equation*}
As a result, we can reproduce the results from previous analyses cited above by simply exploring the different estimations of the quark momentum in $p_\perp$ in the triple product and its asymmetry.

\subsection{Angular observables} \label{subsec : diboson angular observables}

Looking in Table \ref{tab:directobs}, we see two other observables, that are not triple products, already investigated in diboson production : the observable from Ref.~\cite{Azatov:2019xxn} and the signed azimuthal angle difference. 

For the first, we follow the construction of the observable described in Subsection~\ref{subsec:review}.  Namely, the $W$ momentum is needed to reconstruct the scattering plane normal $\Hat{n}_{scat}$ and requires the neutrino momentum $p_\nu$. We replicate the reconstruction method in Ref.~\cite{Azatov:2019xxn} to recover $p_\nu$. Assuming a massless neutrino, its transverse momentum is identified with the missing transverse momentum of the event. The energy of the neutrino momentum is then calculated such that the $W$ boson is on-shell. If two real solutions are possible one of them is randomly chosen or, if no real solution exists, $p_z^{\nu}$ is taken as the value which minimises the electron-neutrino invariant mass. The angles $\phi_W$ and, if applicable, $\phi_Z$ are then derived as in Section~\ref{subsec:review}. From now on, the observable is noted as $\sin{\phi}_{WZ/\gamma}$ with a different definition in each process to ease the notation :
\begin{equation}\label{eq:barducci observable definition}
\begin{split}
    \sin{\phi}_{WZ} & = \sin{2\phi_W}+ \sin{2\phi_Z}, \\
    \sin{\phi}_{W\gamma} & = \sin{2\phi_W}.
\end{split}
\end{equation}
For the second, we look at the electron momentum and the neutral boson momentum.  
Therefore, Eq.~(\ref{eq:genericasym}) dictates that the asymmetries $\Delta \sin{\phi}_{WZ/\gamma}$ and $\Delta \left( \Delta \phi_{eZ/\gamma} \right)$ are respectively defined as
\begin{equation}\label{eq:sinphiVVasym}
    \Delta \sin{\phi}_{WZ/\gamma} = \sigma (\sin{\phi}_{WZ/\gamma}>0) - \sigma (\sin{\phi}_{WZ/\gamma}<0) ,
\end{equation}
where the upper and lower bounds are $\pm$ 1 in this asymmetry, and 
\begin{equation}
    \Delta \left( \Delta \phi_{eZ/\gamma} \right) = \sigma \left( \Delta \phi_{eZ/\gamma}>0 \right) - \sigma \left( \Delta \phi_{eZ/\gamma}<0 \right) ,
\end{equation}
with the upper and lower bounds respectively $\pm \pi$.

%% file: results.tex
\subsection{$WZ$ Results}

We start by checking the interference suppression. The cross-sections for the SM $\sigma(SM)$, for the square of the $\mathcal{O}\left(\Lambda^{-2}\right)$ amplitudes, $\sigma_{\Lambda^{-4}}(\obothi)$, and for the interference $\sigma(\obothi)$ with the two dimension-six operators in $WZ$ production are presented in Table~\ref{tab:xsecWZATLAS}. The interference cross-sections show a suppression by about two orders of magnitude of $\sigma(\obothi)$ compared to the Schwartz bound, \textit{i.e.} twice the geometric mean of $\sigma(SM)$ and $\sigma_{\Lambda^{-4}}(\obothi)$. The absolute value interference cross-section, $\sigma^{|int|}$, and the largest measurable asymmetry, $\sigma^{|meas|}$, are displayed in the same table. They assess that the origin of this suppression is mainly due to the cancellation over the phase space as expected. Additionally, their ratio implies that at most about a third of the interference can be recovered from measured distributions. The large phase space cancellation is also responsible for the poor numerical precision of $\sigma(\obothi)$. All cross-sections, in particular $\sigma(SM)$ and $\sigma_{\Lambda^{-4}}(\obothi)$, are given here at LO and using the same setting as for the interference in order to understand the interference suppression and to check the validity of the scale expansion. $WZ$ production cross-section in the SM is known at NNLO~\cite{Grazzini:2016swo}, and is about a factor 2 bigger than the LO prediction at 13TeV.  The large value of $\sigma^{|meas|}$ for $C_i/\Lambda^2\sim1\,\text{TeV}^{-2}$ compared to $\sigma(SM)$ and the fact that it is quite close to $\sigma^{|int|}$ show that differential distributions could improve significantly the sensitivity to the interference for both operators compared to the total cross-section and that a large part of the phase space cancellation is experimentally accessible. 

\begin{table}[t]
    \centering
    \begin{tabular}{|c|c|c|}
    \hline
          Process & $W^+Z \rightarrow \mu^-\mu^+ e^+ \nu_e$ & $W^-Z \rightarrow \mu^-\mu^+ e^- \Tilde{\nu_e}$  \\
    \hline
          $\sigma(SM)$ & 15.74(2) fb & 9.88(1) fb  \\
    \hline
          $\delta_{PDF}$ & 3.45\% & 3.78\% \\
    \hline
    \hline
          $\sigma(\mathcal{O}_{\widetilde{W}WW})$ & 0.047(4)  fb & -0.033(3) fb \\
    \hline
          Schwartz Bound & 16.13 fb & 8.85 fb \\
    \hline
          $\sigma^{|int|}(\mathcal{O}_{\widetilde{W}WW})$ & 3.302(4) fb &  2.028(3) fb \\
    \hline
          $\sigma^{|meas|}(\mathcal{O}_{\widetilde{W}WW})$ & 1.084(4) fb & 0.634(3) fb \\
    \hline
          $\sigma_{\Lambda^{-4}}(\mathcal{O}_{\widetilde{W}WW})$ & 4.133(5) fb &  1.982(3)  fb \\
    \hline
    \hline
          $\sigma(\mathcal{O}_{\phi\widetilde{W}B})$ & 0.0086(7) fb & -0.0066(4) fb  \\
    \hline
          Schwartz Bound & 1.21 fb & 0.76 fb \\
    \hline
          $\sigma^{|int|}(\mathcal{O}_{\phi\widetilde{W}B})$ & 0.5467(7) fb & 0.3533(4) fb \\
    \hline
          $\sigma^{|meas|}(\mathcal{O}_{\phi\widetilde{W}B})$ & 0.1807(7) fb & 0.1100(4) fb \\
    \hline
          $\sigma_{\Lambda^{-4}}(\mathcal{O}_{\phi\widetilde{W}B})$ & 0.0231(3) fb & 0.0145(2) fb \\
    \hline
    \end{tabular}
    \caption{cross-sections in 2 dileptonic decay channels of $WZ$ production for the ATLAS fiducial phase space at $\sqrt{s}=13$TeV for the SM, the interference with one dimension-six operator, $\sigma(\mathcal{O}_i)$ %\CD{the hat in parentheses does not work in caption?!?} %$\sigma\left(\obothi\right)$
    and for the square of the $\mathcal{O}\left(\Lambda^{-2}\right)$ amplitudes, $\sigma_{\Lambda^{-4}}(\mathcal{O}_i)$. Errors are from the numerical integration and written in the brackets. For the interferences, we also display the absolute value cross-sections $\sigma^{|int|}$ and the measurable absolute value cross-sections $\sigma^{|meas|}$.  $\delta_{PDF}$ represents the uncertainty associated with PDFs taken as the envelope of the replicas for the SM. The Wilson coefficients $C_{\widetilde{W}WW}$ and $C_{\phi\widetilde{W}B}$ are set to 1 and the NP scale $\Lambda$ is 1 TeV but the results can be re-scaled for any other value. %\CD{Espsign in table} \JT{Agree}
    }
    \label{tab:xsecWZATLAS}
\end{table}

Before heading to the triple products with the four possible replacements of the quark momentum, we now turn to the asymmetries of the triple products displayed in Table~\ref{tab:comparison triple products}. The goal is to show the various configurations, measurable and unmeasurable, that we have explored. 
The first line is the ideal triple product with the quark momentum as a reference point. It is compared to triple product with the electron momentum substituted with the muon momentum. We see that the asymmetry of the latter is at least one order below the asymmetry of the former thus disqualifying the muon as the final state fermion to build the best measurable triple product. Then, the best measurable configuration for the triple product is shown: with the longitudinal component of the sum of visible particles as a proxy for the quark momentum, the Z momentum and the electron momentum.
\begin{table}[h!]
    \centering
    \begin{tabular}{c|c|c}
        Triple products configurations & $\mathcal{O}_{\widetilde{W}WW}$ & $\mathcal{O}_{\phi\widetilde{W}B}$ \\
    \hline
        $\left(\vec{p}_q, \vec{p}_Z, \vec{p}_e \right)$ & -1.612(4)  & -0.3888(7) \\
        $\left(\vec{p}_q, \vec{p}_Z, \vec{p}_{\mu^-} \right)$ & -0.184(4)  & -0.0271(7) \\
    \hline
        $\left([0,0,p_\sum^z], \vec{p}_Z, \vec{p}_e \right)$ & -0.628(4)  & -0.1207(7) \\
    \hline
        $\left(\vec{p}_W, \vec{p}_{\mu^-}, \vec{p}_e \right)$ & 0.535(4)  & 0.0965(7) \\
        $\left(\vec{p}_W, \vec{p}_{\mu^+}, \vec{p}_e \right)$ & 0.511(4)  & 0.1009(7) \\
        $\left([0,0,p_W^z], \vec{p}_e, \vec{p}_{\mu^-} \right)$ & -0.227(4)  & -0.0594(7)  \\
        $\left(\vec{p}_W, \vec{p}_{\mu^-}, \vec{p}_{\mu^+} \right)$ & -0.080(4)  & -0.0110(7) \\
        $\left([0,0,p_\sum^z], \vec{p}_W, \vec{p}_Z \right)$ & -0.045(4)  & -0.0086(7) \\
        $\left([0,0,p_e^z], \vec{p}_{\mu^-}, \vec{p}_W \right)$ & 0.028(4)  & 0.0061(7) \\
    \hline
        $\left(\vec{p}_e, \vec{p}_{\mu^-}, \vec{p}_{\mu^+} \right)$ & -0.025(4)  & -0.004(7) \\
        $\left([0,0,p_e^z], \vec{p}_{\mu^-}, \vec{p}_{\mu^+} \right)$ & -0.029(4)  & -0.0061(7) \\
        $\left([0,0,p_{\mu^-}^z], \vec{p}_e, \vec{p}_{\mu^+} \right)$ & -0.213(4)  & -0.0244(7) \\
        $\left([0,0,p_{\mu^+}^z], \vec{p}_{\mu^-}, \vec{p}_e \right)$ & 0.252(4)  & 0.0327(7) \\
        $\left([0,0,p_{\sum}^z], \vec{p}_e+\vec{p}_{\mu^-}, \vec{p}_{\mu^+} \right)$ & -0.362(4) & -0.0582(7) \\
        $\left([0,0,p_{\sum}^z], \vec{p}_e+\vec{p}_{\mu^+}, \vec{p}_{\mu^-} \right)$ & -0.300(4)  & -0.0481(7) \\
        $\left([0,0,p_{\sum}^z], \vec{p}_e-\vec{p}_{\mu^-}, \vec{p}_{\mu^+} \right)$ & -0.047(4)  & -0.0097(7) \\
        $\left([0,0,p_{\sum}^z], \vec{p}_e-\vec{p}_{\mu^+}, \vec{p}_{\mu^-} \right)$ & -0.160(4)  & -0.0279(7) 
    \end{tabular}
    \caption{Table of the asymmetries measured with respect to different configurations of the triple product observable in $W^+Z$ production. The first column displays the momenta chosen to build the observable and the last two columns show the asymmetries in fb obtained for each operator%, respectively $\mathcal{O}_{\widetilde{W}WW}$ and $\mathcal{O}_{\phi\widetilde{W}B}$
    . The numerical errors are put in brackets.}
    \label{tab:comparison triple products}
\end{table}
After that, using the leptons with W turns out quite close to the best measurable configuration ($\sim$85\% of the best configuration). However, we cannot measure the W momentum and any approximation would reduce further the asymmetry. Several combinations of the final three leptons are investigated. We are forced the use the longitudinal component of one lepton as the triple product using the three complete momenta almost cancels. By taking the sum of the electron and either the muon or anti-muon we can approach 50\% of the best configuration. The difference does not improve the asymmetries either. 

Henceforth the rest of the triple products are only built with the electron and the Z momenta and the three aforementioned approximations for the quark momenta: the longitudinal component of the electron, of the Z boson and of the sum over all the final visible lepton. The largest asymmetries are obtained with the latest for both operators and for both channels as shown in Table~\ref{tab:WZasymmetryobservablesATLAS}.

\begin{table}[h!]
    \centering
    \begin{tabular}{|c|c|c|c|}
    \hline 
        Process & \multicolumn{3}{c|}{$W^+ Z \rightarrow \mu^- \mu^+ e^+ \nu_e $} \\
    \hline
        Operators & $SM$ & $\mathcal{O}_{\widetilde{W}WW}$ & $\mathcal{O}_{\phi\widetilde{W}B}$  \\
    \hline
       $\Delta p_\perp (p_e,p_q)$ & -0.04(2) & -1.612(4)  & -0.3888(7)  \\
    \hline
       $\Delta p_\perp (p_e,p_\sum^z)$  & -0.02(2) & -0.628(4)  & -0.1207(7)   \\
    \hline
       $\Delta p_\perp (p_e,p_e^z)$ & 0.0(2)  & -0.535(4) &  -0.1173(7) \\
    \hline
       $\Delta p_\perp (p_e,p_Z^z)$ & -0.01(2)  & -0.527(4) &  -0.0874(7) \\
    \hline
       $\Delta \sin{\phi_{WZ}}$ & -0.03(2)  & -0.321(4) &  0.0031(7)  \\
    \hline
       $\Delta \left( \Delta \phi_{eZ} \right)$ & 0.07(2)  & 0.196(4) &  0.0688(7)  \\
    \hline
      SM stat err $30~\text{fb}^{-1}$ & \multicolumn{3}{c|}{0.7}  \\
    \hline
      SM stat err $100~\text{fb}^{-1}$ & \multicolumn{3}{c|}{0.4}  \\
    \hline
      SM stat err $3000~\text{fb}^{-1}$ & \multicolumn{3}{c|}{0.07}  \\
    \hline
    \hline 
        Process & \multicolumn{3}{c|}{$W^- Z \rightarrow \mu^- \mu^+ e^- \Tilde{\nu}_e $} \\
    \hline
        Operators & $SM$ & $\mathcal{O}_{\widetilde{W}WW}$ & $\mathcal{O}_{\phi\widetilde{W}B}$  \\
    \hline
       $\Delta p_\perp (p_e,p_q)$ & -0.08(1) & 1.006(3) & 0.2522(4)  \\
    \hline
       $\Delta p_\perp (p_e,p_\sum^z)$  & -0.03(1) & -0.331(3) & 0.0810(4) \\
    \hline
       $\Delta p_\perp (p_e,p_e^z)$ & -0.01(1) & 0.295(3) & 0.0514(4) \\
    \hline
       $\Delta p_\perp (p_e,p_Z^z)$ & 0.00(1) & 0.295(3) & 0.0627(4)  \\
    \hline
       $\Delta \sin{\phi_{WZ}}$ & -0.02(1) & -0.190(3) & 0.0013(4)   \\
    \hline
       $\Delta \left( \Delta \phi_{eZ} \right)$ & -0.05(1) & 0.022(3) &  0.0109(4) \\
    \hline
      SM stat err $30~\text{fb}^{-1}$ & \multicolumn{3}{c|}{0.6}  \\
    \hline
      SM stat err $100~\text{fb}^{-1}$ & \multicolumn{3}{c|}{0.3}   \\
    \hline
      SM stat err $3000~\text{fb}^{-1}$ & \multicolumn{3}{c|}{0.06} \\
    \hline
    \end{tabular}
    \caption{Asymmetries in fb in the $WZ \rightarrow \mu^-\mu^+ e^+ \nu_e$ and $WZ \rightarrow \mu^-\mu^+ e^- \widetilde{\nu}_e$ channels by using different reference axes for the triple product and $\sin{\phi_{WZ}}$, in the ATLAS fiducial phase space at $\sqrt{s}=13$TeV at the LHC. The statistical errors are displayed using the LO SM cross-sections and several integrated luminosities. }
    \label{tab:WZasymmetryobservablesATLAS}
\end{table}

As mentioned in Subsection \ref{subsec : diboson angular observables}, our generic triple product asymmetry is also compared to the asymmetry of $\sin{\phi_{WZ}}$ from Ref.~\cite{Azatov:2019xxn} and of the signed azimuthal angle difference which has also been studied in $WZ$ and $W\gamma$ production. The values of the asymmetries are presented in Table~\ref{tab:WZasymmetryobservablesATLAS} as well. 

The asymmetry in $p_\perp (p_e, p_\sum)$ for $\mathcal{O}_{\widetilde{W}WW}$ is a factor 2 bigger than the one obtained from $\sin\phi_{WZ}$. $\sin\phi_{WZ}$ is also almost insensitive to the $\mathcal{O}_{\phi\widetilde{W}B}$ contribution. Therefore, the two operators can be distinguished with this channel only by measuring the asymmetries in $p_\perp (p_e, p_\sum)$ and $\sin\phi_{WZ}$. On the contrary, $\Delta \phi_{eZ}$ is less sensitive to $\OWWW$ than $\OBW$ effects in $W^+Z$ and therefore probes other combinations of the operators.  In particular, $\Delta \phi_{eZ}$ is almost blind to $\OWWW$ in $W^-Z$. In the end, the triple product asymmetries with the longitudinal component of sum of the visible as reference axis display better sensitivities.  

By comparing with $\sigma^{|meas|}$, we see that the efficiency of the best triple product is about 50\% and 70\% for the operators $\mathcal{O}_{\widetilde{W}WW}$ and $\mathcal{O}_{\phi\widetilde{W}B}$ respectively. Those observables are not purely CP-violating since the processes are not C-even and taking the sum of the two final state is not enough as also the initial state is not C-even at the LHC. Therefore, the SM contribution (effects from the CKM phase are negligible) to those asymmetries does not have to vanish. However, we found that the asymmetries were well below the percent level and consistent with zero with our numerical precision. This would have to be checked in higher order computations. Similarly, the asymmetries from the square of the $\mathcal{O}(\Lambda^{-2})$ amplitudes are consistent with zero, so they are not shown in Table \ref{tab:WZasymmetryobservablesATLAS}. Overall $\mathcal{O}_{\widetilde{W}WW}$ produces larger CP violating effects than $\mathcal{O}_{\phi\widetilde{W}B}$. However, the normalisation of the operators is arbitrary as long as no UV complete model is introduced.

\begin{figure}[p]
\centering
%\subfloat{
\includegraphics[width=0.47\textwidth,trim={0 0 20pt 20pt} ,clip]{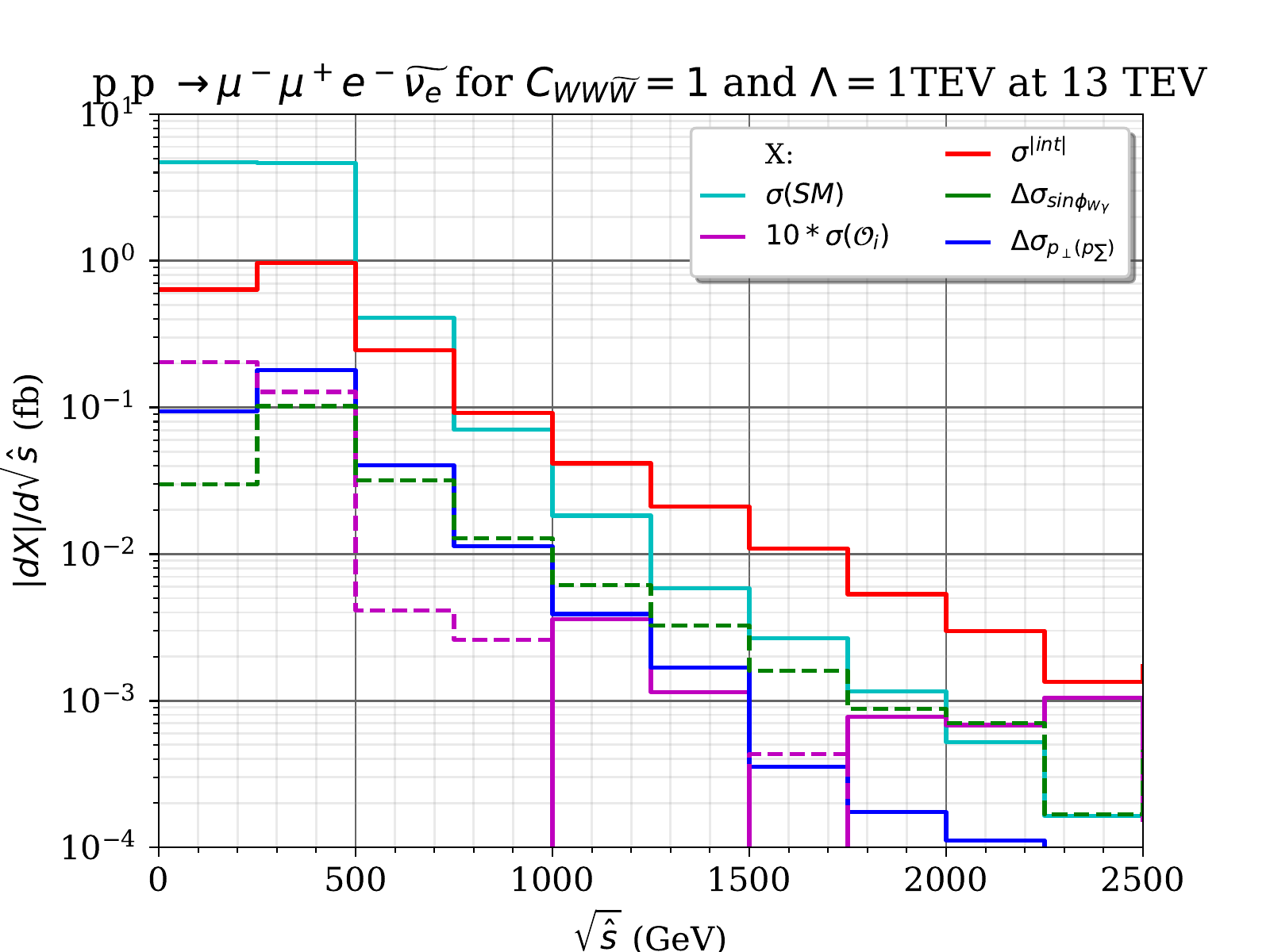}
%}
\quad
%\subfloat{
\includegraphics[width=0.47\textwidth,trim={0 0 20pt 20pt} ,clip]{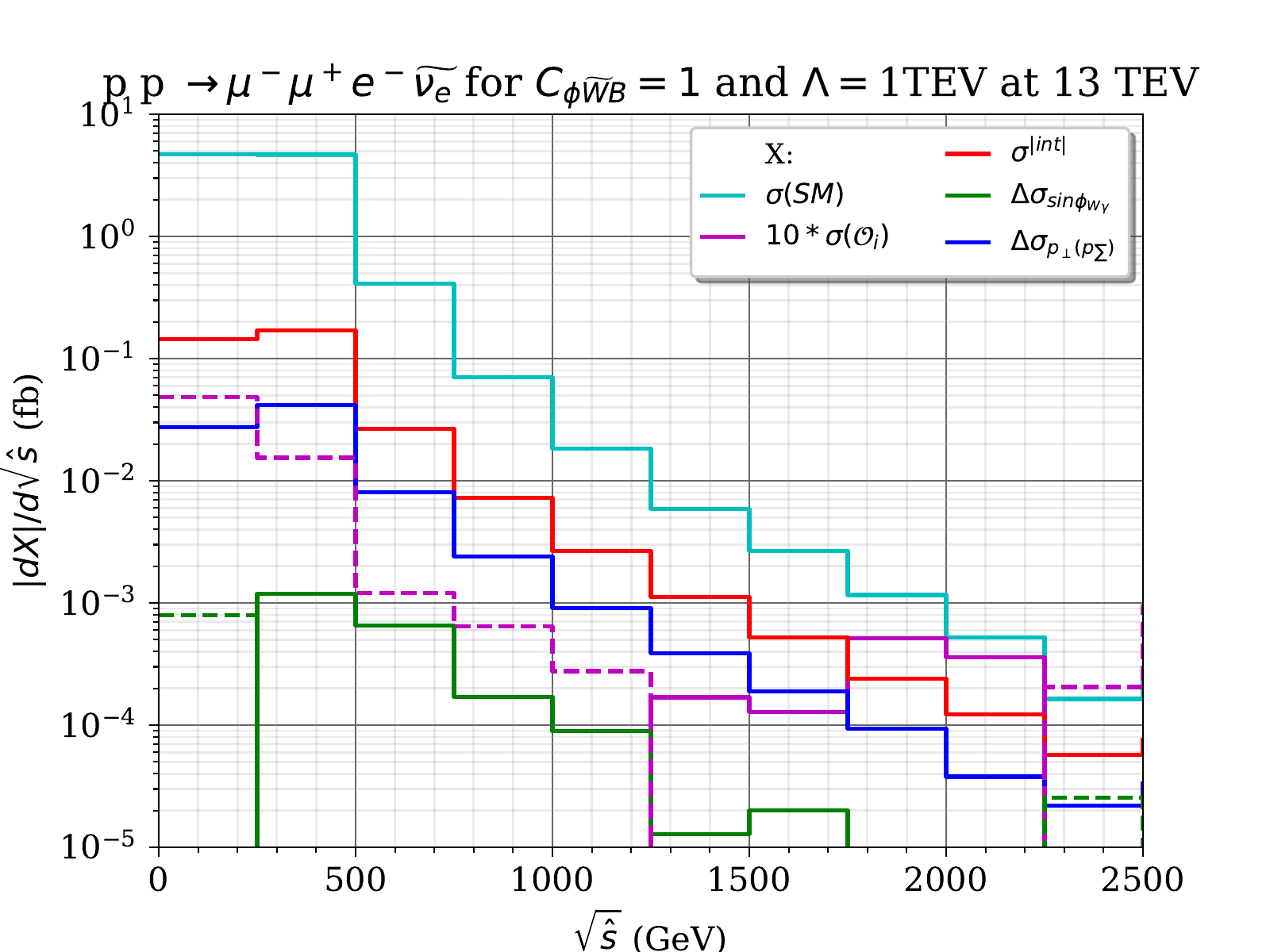}
%}
%\subfloat{
\includegraphics[width=0.47\textwidth,trim={0 0 20pt 20pt} ,clip]{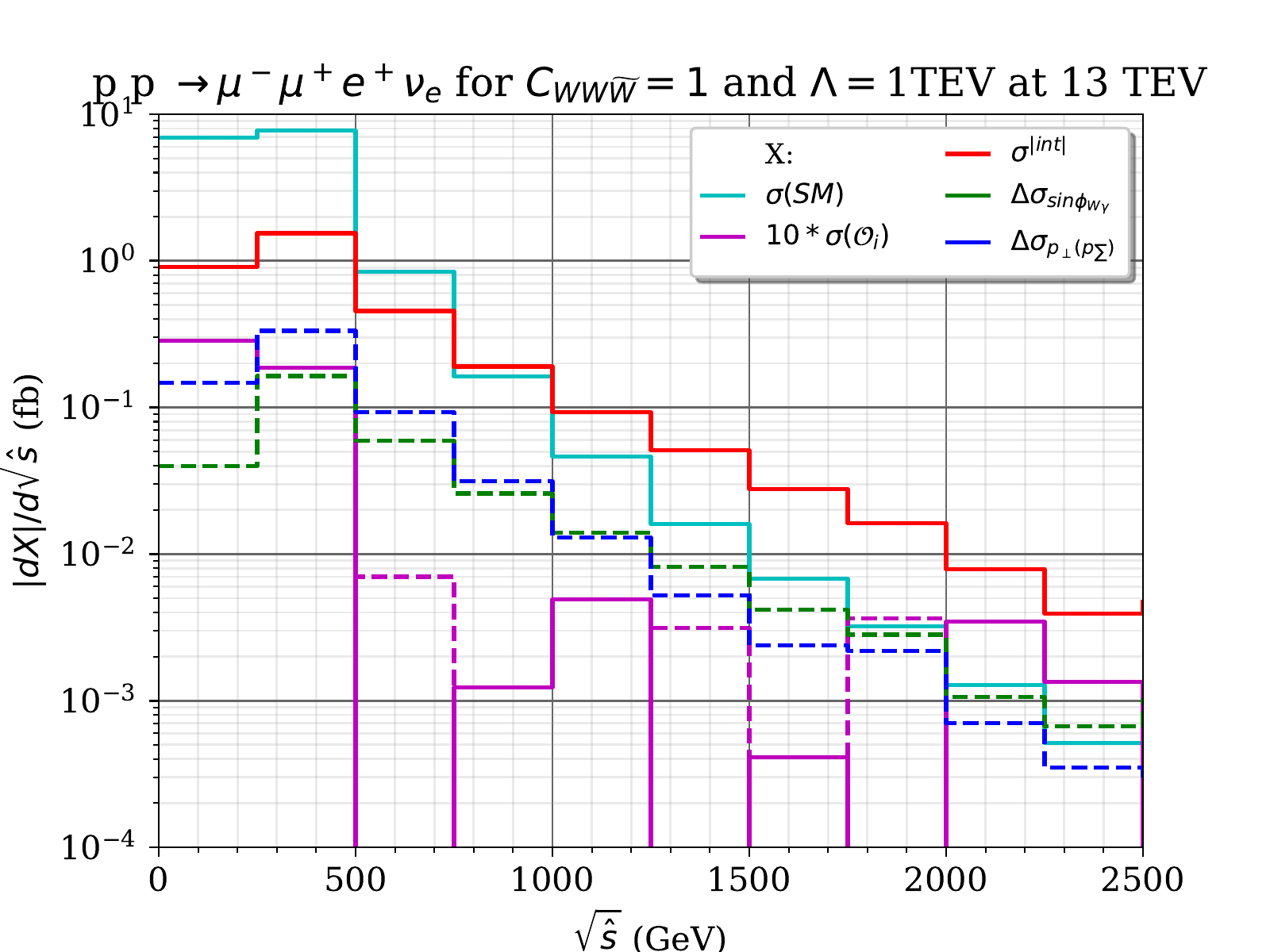}
%}
\quad
%\subfloat{
\includegraphics[width=0.47\textwidth,trim={0 0 20pt 20pt} ,clip]{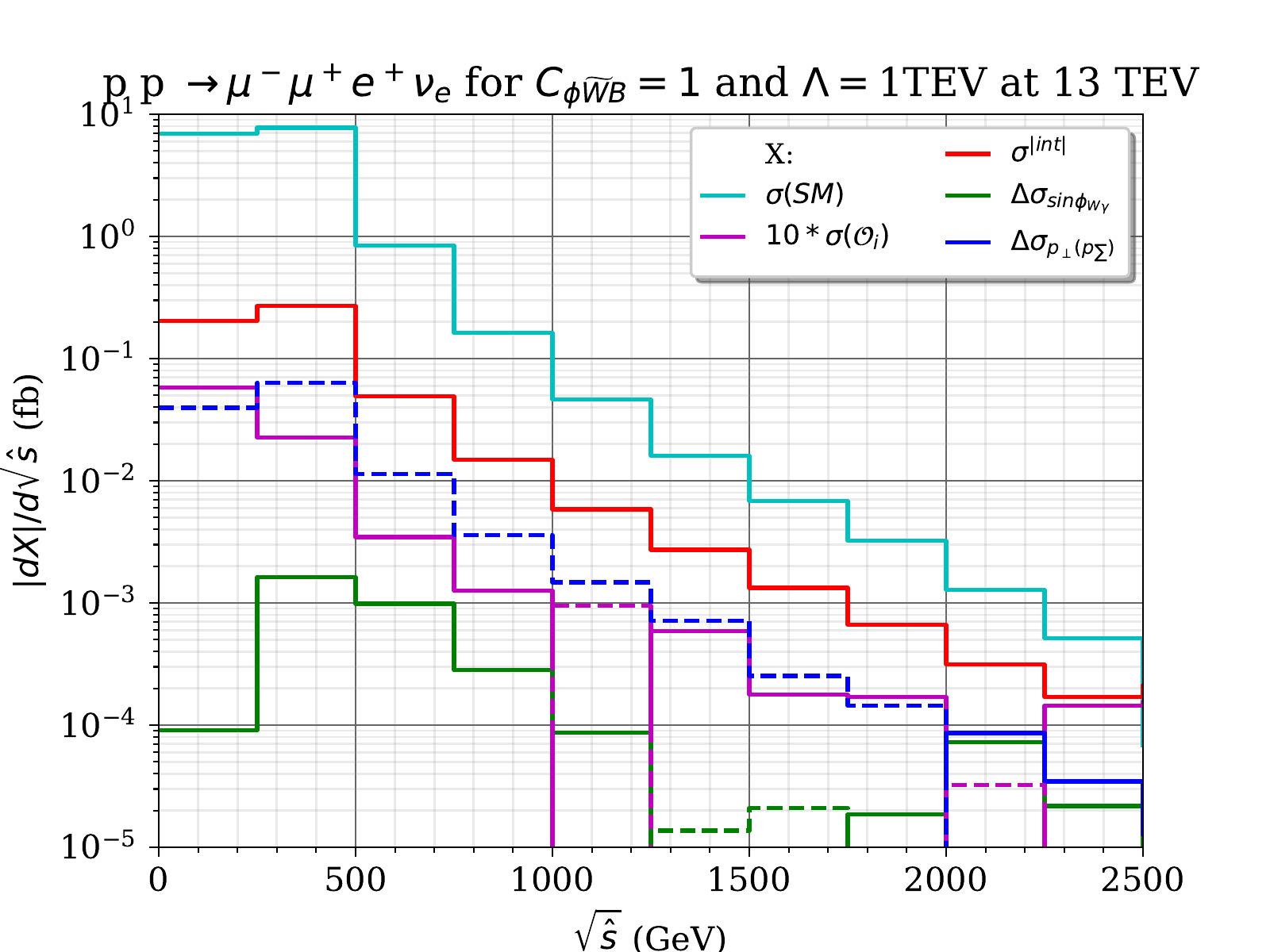}
%}
\caption{Differential cross-sections of SM and interferences with respect to $\sqrt{\Hat{s}}$ in $WZ$ production in the ATLAS fiducial phase space are displayed in light blue and purple respectively while we represent the asymmetries following the true matrix elements in red. The differential triple product asymmetries are drawn in blue and the differential asymmetries for $\sin\varphi_{WZ}$ in green. For $\OBW$, these asymmetries are too small and dominated by numerical errors. The dashed lines correspond to negative values. %\CD{EPSsign : change sign all purple lines(int x-sect.)}
}
\label{fig:ECMZW}
\end{figure}

Finally, we check the energy dependence of the asymmetries and the interference in general. The differential distributions as a function of the center of mass energy, $\sqrt{\hat{s}}$, are displayed in Fig.~\ref{fig:ECMZW}. Positive values are shown with straight lines and negative ones with dashed lines. The interference cross-section is multiplied by 10 to appear alongside the different asymmetry distributions. We see that the interference cross-sections change sign after the first few bins showing they are quickly dominated by numerical fluctuations. We see that those $\sigma^{|int|}$ distributions have harder slopes than the SM as expected from higher dimensional operators contributions especially for $\OWWW$ due to the higher power of momenta in its vertices. We also see that our asymmetry is more effective close to threshold while the one based on $\sin{\phi_{WZ}}$  wins at high energy for $\OWWW$ as expected from its construction. This further demonstrates the complementarity of the two observables.

\subsection{$W\gamma$ Results}

Similarly to $WZ$, we display our results for $W\gamma$ production obtained for the cross-sections and the asymmetries in Table~\ref{tab:xsecWAATLAS}. The $\OWWW$ interference contribution is much smaller compared to the SM even if we consider the absolute value cross-section but this can be understand by the lower energies probed by this process and therefore a harder cut could improve the new physics contribution relatively to the SM. As a result, the contributions from both operators are this time comparable and therefore suggest already that the combination of the two processes can be used to distinguish them. The ratios between the interference cross-section and $\sigma^{|int|}$ are not as large but still signal the presence of strong phase space suppression. Additionally, the ratios between $\sigma^{|meas|}$ and $\sigma^{|int|}$  are smaller. This may be partially due to quark and anti-quark PDFs and the unmeasured photon helicity. If $W$ decays into a neutrino and a muon instead of an electron, the last cut in $W\gamma$ from Section \ref{sec:analysis} would no longer apply but we do not expect the results to vary much. 

\begin{table}[t]
    \centering
    \begin{tabular}{|c|c|c|}
    \hline
          Process & $W^+ \gamma \rightarrow \gamma e^+ \nu_e$ & $W^- \gamma \rightarrow \gamma e^- \Tilde{\nu_e}$  \\
    \hline
          $\sigma(SM)$ & 715.1(8) fb & 589.1(7) fb  \\
    \hline
          $\delta_{PDF}$ & 2.99\% & 3.43\% \\
    \hline
    \hline
          $\sigma(\mathcal{O}_{\widetilde{W}WW})$ & -2.07(4) fb & 1.61(6) fb \\
    \hline
          Schwartz Bound & 337.3 fb & 209.0 fb \\
    \hline
          $\sigma^{|int|}(\mathcal{O}_{\widetilde{W}WW})$ & 33.83(4) fb & 24.76(6) fb \\
    \hline
          $\sigma^{|meas|}(\mathcal{O}_{\widetilde{W}WW})$ & 6.07(4) fb & 6.57(6) fb \\
    \hline
          $\sigma_{\Lambda^{-4}}\left(\OWWW\right)$ & 39.78(5) fb & 18.54(6) fb \\
    \hline
    \hline
          $\sigma(\mathcal{O}_{\phi\widetilde{W}B})$ & 2.75(4) fb & -2.09(3) fb  \\
    \hline
          Schwartz Bound & 96.3 fb & 82.4 fb \\
    \hline
          $\sigma^{|int|}(\mathcal{O}_{\phi\widetilde{W}B})$ & 34.00(4) fb & 26.37(3)  fb \\
    \hline
          $ \sigma^{|meas|}(\mathcal{O}_{\phi\widetilde{W}B})$ & 9.43(4) fb & 9.53(3) fb \\
    \hline
          $\sigma_{\Lambda^{-4}}\left(\OBW\right)$ & 3.239(4) fb & 2.878(3) fb \\
    \hline
    \end{tabular}
    \caption{Cross-sections in the dileptonic decay channel of $W\gamma$ production for the ATLAS fiducial phase space at $\sqrt{s}=13$TeV for the SM, the interference with each dimension-six operator, $\sigma(\mathcal{O}_i)$, and for the square of the $\mathcal{O}\left(\Lambda^{-2}\right)$ amplitudes , $\sigma_{\Lambda^{-4}}(\mathcal{O}_i)$. Errors are from the numerical integration. For the interferences, we also display the absolute value cross-sections $\sigma^{|int|}$ and the measurable absolute value cross-section $\sigma^{|meas|}$. The Wilson coefficients $C_{\widetilde{W}WW}$ and $C_{\phi\widetilde{W}B}$ are set to 1 and the NP scale $\Lambda$ is 1 TeV but the results can be re-scaled for any other value.}
    \label{tab:xsecWAATLAS}
\end{table}

\begin{table}[h!]
    \centering
    \begin{tabular}{|c|c|c|c|c|c|}
    \hline 
        Process & \multicolumn{3}{c|}{$W^+\gamma \rightarrow \gamma e^+ \nu_e $} \\
    \hline
        Operators & $SM$ & $\mathcal{O}_{\widetilde{W}WW}$ & $\mathcal{O}_{\phi\widetilde{W}B}$  \\
    \hline
       $\Delta p_\perp (p_e,p_q)$ & 7.7(8) & -13.81(4) & 22.23(4)  \\
    \hline
       $\Delta p_\perp (p_e,p_\sum^z)$ & 0.8(8) & -4.60(4) & 5.59(4)   \\
    \hline
       $\Delta p_\perp (p_e,p_\gamma^z)$ & 0.5(8) & -5.62(4) & 7.59(4)   \\
    \hline
       $\Delta p_\perp (p_e,p_e^z)$ & 0.6(8) & 1.11(4) & 0.42(4) \\
    \hline
       $\Delta \sin{\phi_{W\gamma}}$ & -0.1(8) & -0.31(4) & -0.79(4)  \\
    \hline
       $\Delta \left( \Delta \phi_{eZ} \right)$ & -4.5(8) & -5.85(4) & 7.16(4) \\
    \hline
      SM stat err $30~\text{fb}^{-1}$ & \multicolumn{3}{c|}{4}  \\
    \hline
      SM stat err $100~\text{fb}^{-1}$ & \multicolumn{3}{c|}{2}  \\
    \hline
      SM stat err $3000~\text{fb}^{-1}$ & \multicolumn{3}{c|}{0.4}  \\
    \hline
    \hline 
        Process & \multicolumn{3}{c|}{$W^-\gamma \rightarrow \gamma e^- \Tilde{\nu}_e $} \\
    \hline
        Operators & $\Delta\sigma(SM)$ & $\Delta\sigma(\mathcal{O}_{\widetilde{W}WW})$ & $\Delta\sigma(\mathcal{O}_{\phi\widetilde{W}B})$ \\
    \hline
       $\Delta p_\perp (p_e,p_q)$ & 5.3(7) & 10.65(3) & -17.27(3) \\
    \hline
       $\Delta p_\perp (p_e,p_\sum^z)$ & 1.2(7) & 2.34(3) & -4.15(3) \\
    \hline
       $\Delta p_\perp (p_e,p_\gamma^z)$ & 0.1(7) & -1.68(3) & 1.48(3) \\
    \hline
       $\Delta p_\perp (p_e,p_e^z)$ & 0.9(7) & 5.09(3) & -7.07(3) \\
    \hline
       $\Delta \sin{\phi_{W\gamma}}$ & -0.4(7) & -1.87(3) & 1.22(3) \\
    \hline
       $\Delta \left( \Delta \phi_{eZ} \right)$ & 1.2(7) & -6.17(3) & 8.46(3)   \\
    \hline
      SM stat err $30~\text{fb}^{-1}$ & \multicolumn{3}{c|}{4}  \\
    \hline
      SM stat err $100~\text{fb}^{-1}$ & \multicolumn{3}{c|}{2}  \\
    \hline
      SM stat err $3000~\text{fb}^{-1}$ & \multicolumn{3}{c|}{0.4} \\
    \hline
    \end{tabular}
    \caption{Asymmetries in fb in the $W \gamma \rightarrow \gamma e^+ \nu_e$ and $W\gamma \rightarrow \gamma e^- \widetilde{\nu}_e$ channels by using different reference axes for the four observables, the triple product, the two triple products with the beam direction correction factor and the Barducci observable, in the ATLAS fiducial phase space at $\sqrt{s}=13$TeV at the LHC. The statistical errors are displayed using the SM cross-sections and several integrated luminosities.     }
    \label{tab:WAasymmetryobservables}
\end{table}

In $W\gamma$ production the triple product can only be constructed with the electron momentum and the photon momentum. Therefore, we only have to look for the best proxy for the quark momentum as the same argument about the non-measurability of the quark momentum stands. As in the $WZ$ production, we consider the longitudinal component of the electron, of the photon and of the sum over all the final visible particles as shown in Table~\ref{tab:WAasymmetryobservables}.
We first check the triple product asymmetries with the quark momentum. This particular triple product produces asymmetries around 40\% of $\sigma^{|int|}$ for $\mathcal{O}_{\widetilde{W}WW}$ and 65\% for $\mathcal{O}_{\phi\widetilde{W}B}$. The asymmetry in the triple product with the quark momentum is a bit less effective in $W\gamma$ production than in $WZ$ production but remains quite good. One drawback is that the SM asymmetry is more important. 
The largest asymmetry for $W^+\gamma$ production is obtained with the photon longitudinal momenta while the electron longitudinal momentum give the largest asymmetry for $W^-\gamma$. Those asymmetries have quite large efficiencies as their ratios with $\sigma^{|meas|}$ around 80\% and even above to 90\% for $\OWWW$ in $W^+\gamma$. 
Those large efficiencies are in contrast with the poor sensitivity obtained in this channel for the observable proposed in Ref~\cite{Azatov:2019xxn}, especially for $\OWWW$. Unlike in $WZ$, $\sin{\phi_{W\gamma}}$ appears ineffective in tracking CP-violating operators in $W\gamma$ while $\Delta \phi_{eZ}$ provides results almost similar or  even better than the best triple product in $W^+\gamma$ depending on the operator and the sign of the W boson. By taking the best observable for each channel and each operator, the efficiencies are all very close to 90\% except for $\OBW$ in $W^+\gamma$ which is at 80\%.

The asymmetries derived from the squared amplitudes are consistent with zero and always at least more than one order of magnitude below those of the interferences.

\begin{figure}[p]
\centering
%\subfloat{
\includegraphics[width=0.47\textwidth,trim={0 0 30pt 20pt} ,clip]{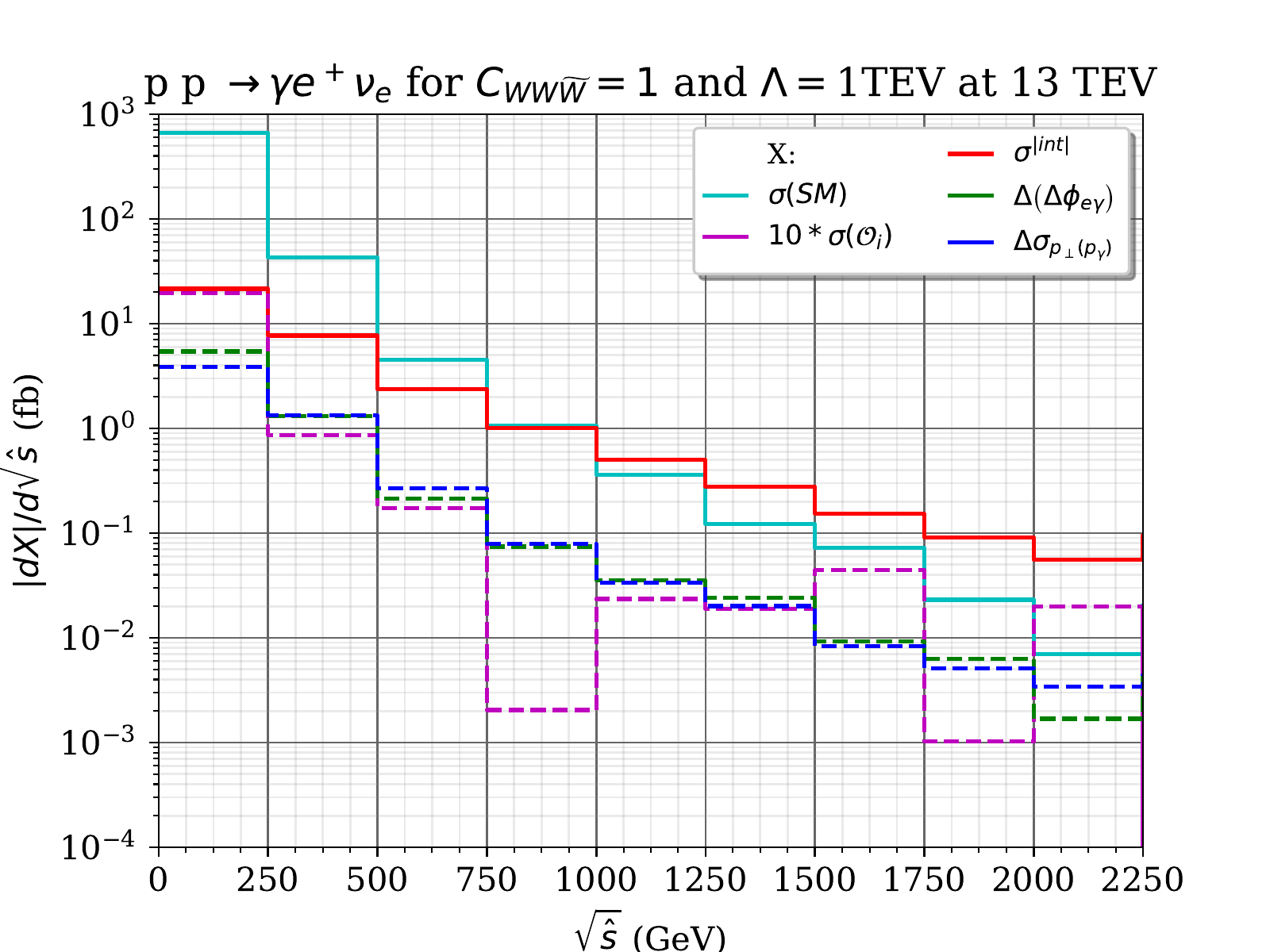}
%}
\quad
%\subfloat{
\includegraphics[width=0.47\textwidth,trim={0 0 30pt 20pt} ,clip]{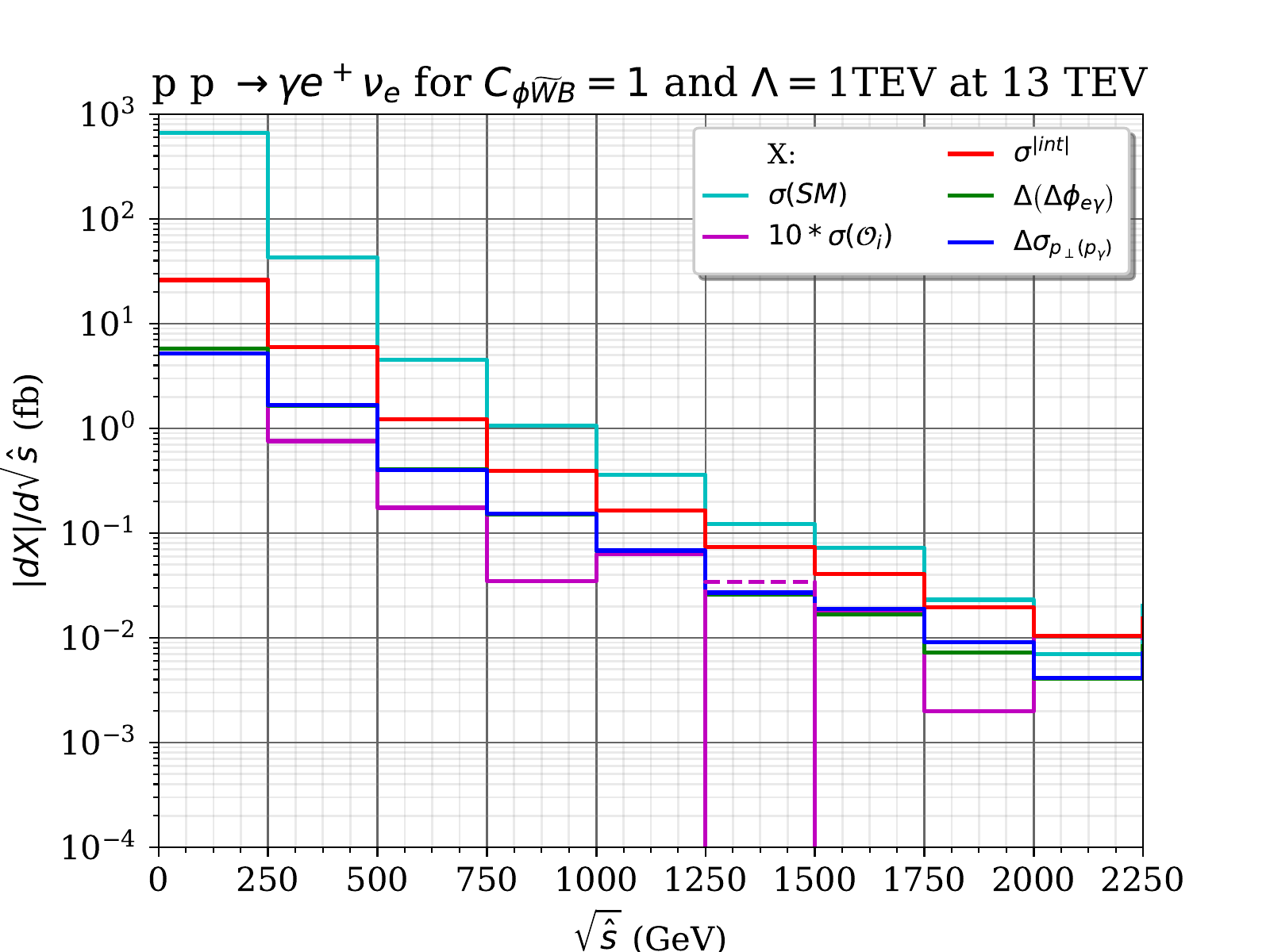}
%}
%\subfloat{
\includegraphics[width=0.47\textwidth,trim={0 0 30pt 20pt} ,clip]{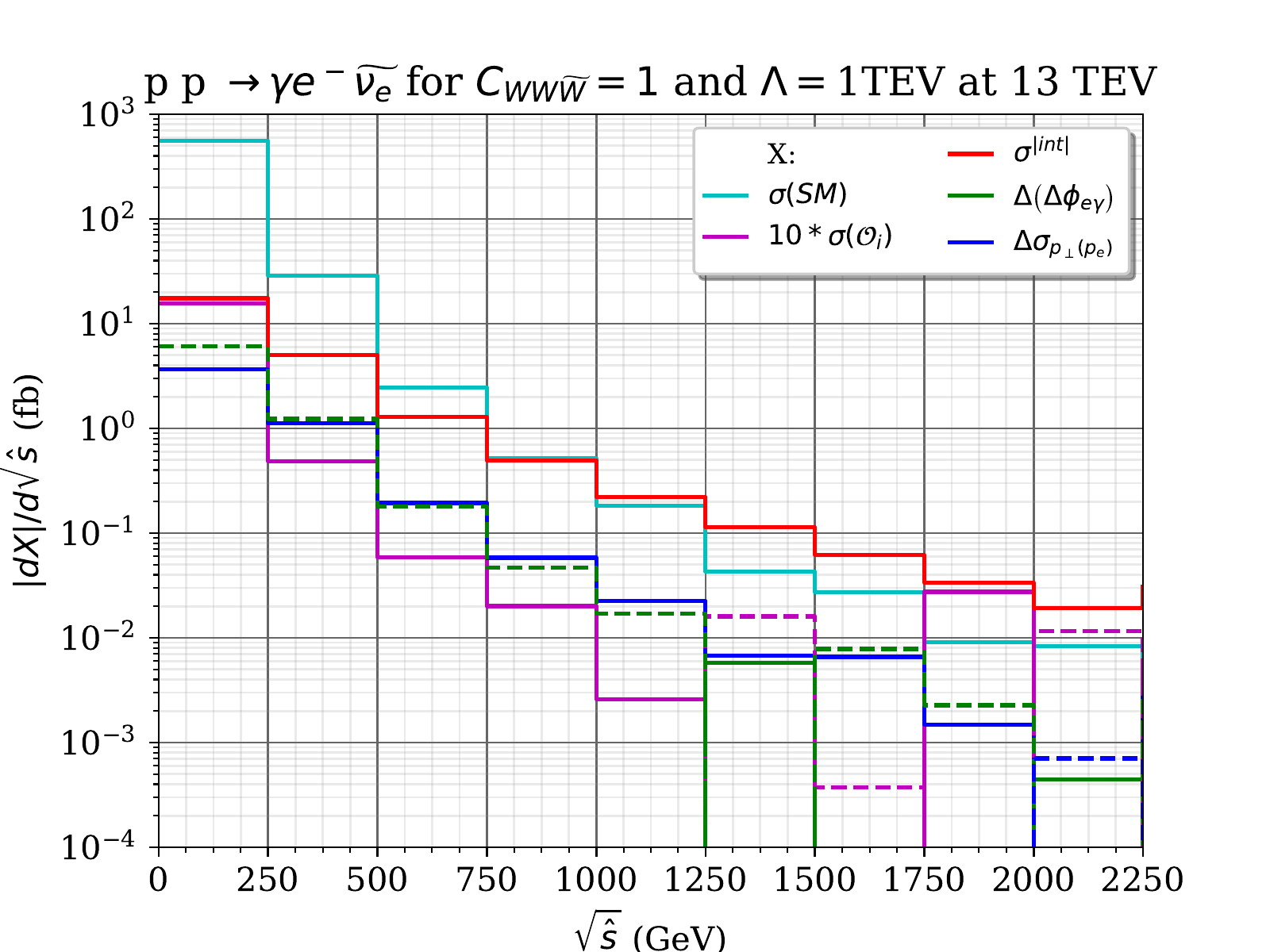}
%}
\quad
%\subfloat{
\includegraphics[width=0.47\textwidth,trim={0 0 30pt 20pt} ,clip]{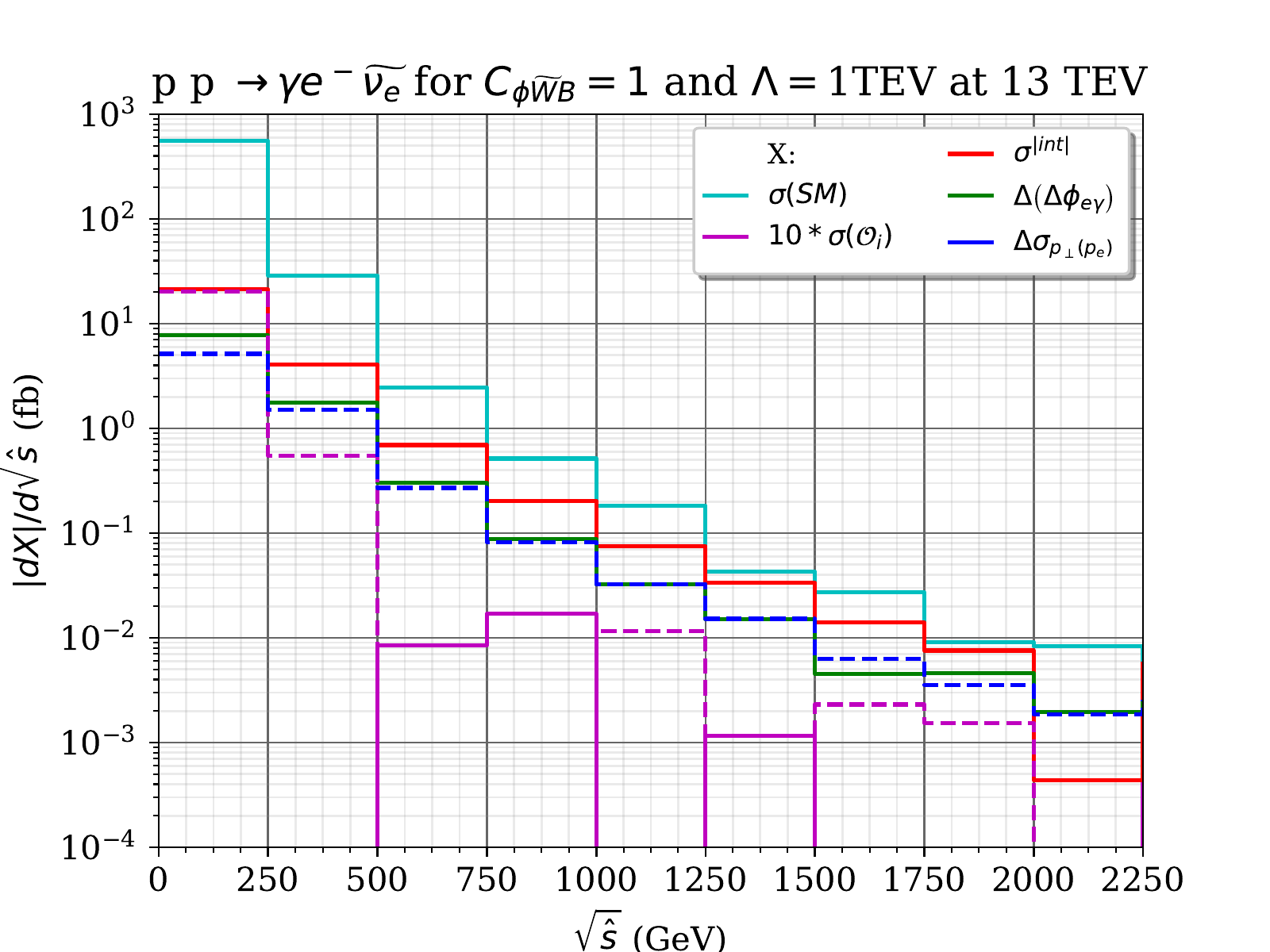}
%}
\caption{Differential cross-sections of SM and interferences with respect to $E_{c.o.m.}$ in $W\gamma$ production in the ATLAS fiducial phase space are displayed in light blue and purple respectively while we represent the asymmetries following the true matrix elements in red. The differential triple product asymmetries are drawn in blue and the differential asymmetries of $\sin\varphi_{W\gamma}$ in green. The bins in dashed lines correspond to negative values. %\CD{I would remove sinphiwa in both channel and put delta phi and the best triple product} \JT{Done}
}
\label{fig:EcmAW}
\end{figure}

Finally, the distributions of the asymmetries as a function of $\sqrt{\Hat{s}}$ are displayed in Fig.~\ref{fig:EcmAW}. The interference cross-sections are again multiplied by 10, the distributions have the same colours as in Figure~\ref{fig:ECMZW} and show a similar behaviour as for $WZ$ production: interference cross-sections begin to fluctuate in high-energy bins and even change sign, the $\sigma^{|int|}$ asymmetries have a harder slope than the SM and the $p_\perp(p_e, p_\sum)$ asymmetry follows quite closely $\sigma^{|int|}$. The distributions of $\Delta \phi_{e\gamma}$ follow closely the triple product ones over the whole energy range irrespective of the channel or the operator.

\subsection{Sensitivity}

As our aim here has always been to motivate observables that could be used in future more detailed analyses to draw the most stringent constraints operators, we follow the method described in Ref.\cite{Kumar:2008ng} to estimate the sensitivities to the two operators. For those estimates, we assume that the SM process is the only background and we use the LO SM cross-sections as the total measured cross-sections. We also assume the SM to be symmetric, which does not have to be the case even without CP violation as our observables are not pure CP-odd. The sensitivities are obtained by fixing the signal over background ratio $(S/\sqrt{B} = 2) \sim 2 \sigma$.
Firstly, we compare our result for $p_\perp (p_e, p_Z)$, which is the variable they use in Ref.\cite{Kumar:2008ng}, to their result.
The result presented, $\Delta \sigma = \Tilde{\lambda}\times (3\times 10^3~ \text{fb})$, actually considers a coupling constant $\Tilde{\lambda}_Z$ in an anomalous interaction rather than a Wilson coefficient form an EFT. Using $\Tilde{\lambda}_Z = 6  C_{\widetilde{W}WW} M_W^2 / g \Lambda^2$ from Ref.~\cite{Bohm:1987ck}, we are able to translate the result such that
\begin{equation*}
    \Delta \sigma = C_{\widetilde{W}WW} \times (1.8 \times 10^2)~\text{fb}.
\end{equation*} 
Then, we apply the decay fractions of the $W$ and $Z$ bosons into a single leptonic channel as provided by the PDG, which are $10.7\%$ and $3.37\%$ respectively, to match the final states between our analysis and Ref.\cite{Kumar:2008ng}. This results in $\Delta \sigma = C_{\widetilde{W}WW} \times 0.644~\text{fb}$. 
By adding our values of $p_\perp(p_e, p_Z)$ in both $W^\pm Z$, we get $\Delta \sigma = C_{\widetilde{W}WW} \times 0.798~\text{fb}$. There is a small difference which can be understood mainly from the different cuts. 

\begin{figure}[h!]
    \centering
    \includegraphics[scale=0.9]{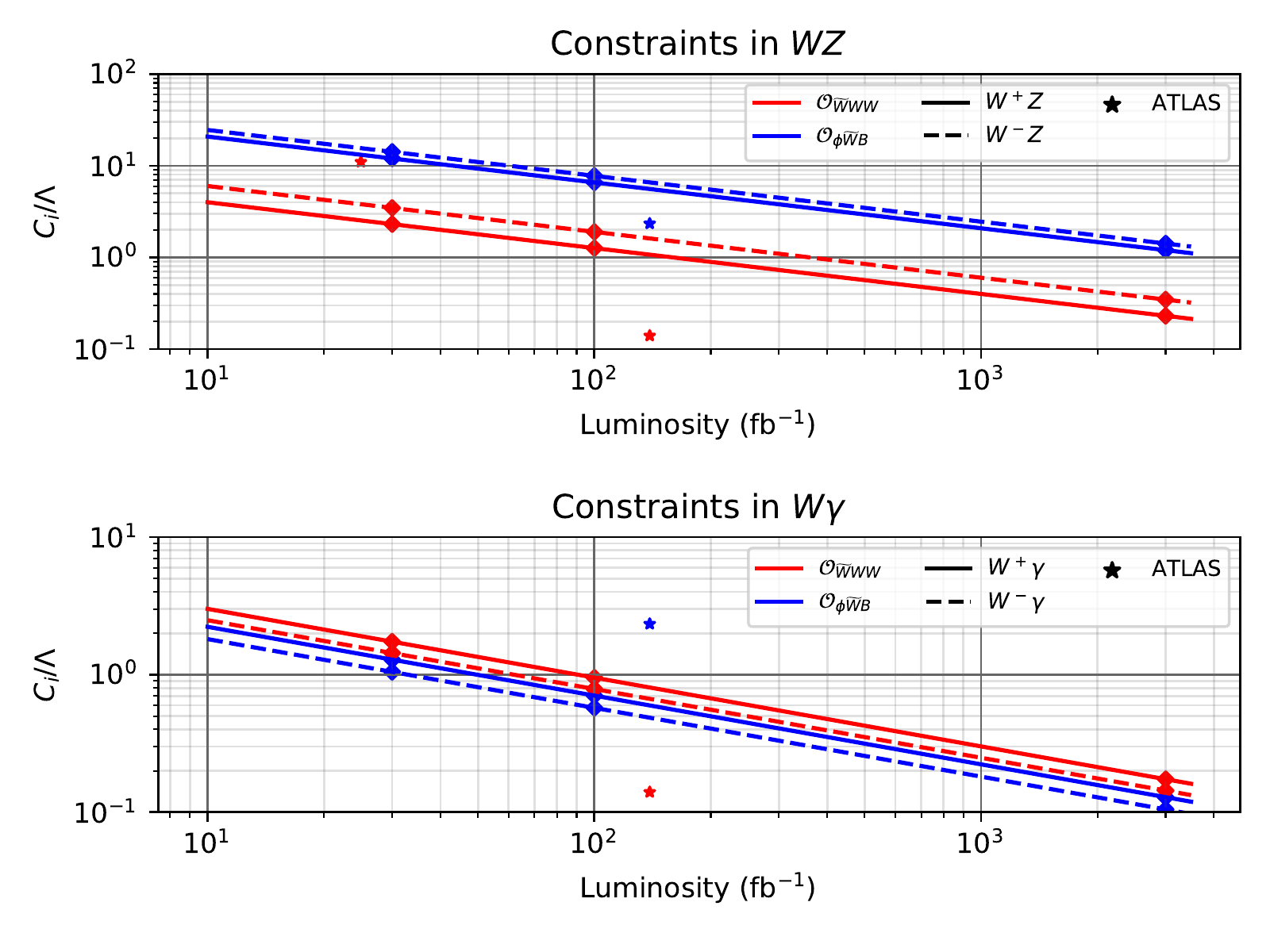}
    \caption{Sensitivity at 95\%CL as a function of the integrated luminosity for each process and each operator using the best observable in each case and the SM at LO as the only background and assuming it to be symmetric. %\JT{$\Delta \phi_{e\gamma}$ used for $W^-\gamma$} 
    }
    \label{fig:wilson coeff constr}
\end{figure}

Secondly, we apply this method with $p_\perp(p_e, p_\sum)$ for the $WZ$ channels, with $p_\perp(p_e, p_\gamma)$ for $W^+\gamma$, and with $\Delta \phi_{e\gamma}$ for $W^-\gamma$ and for different luminosities. They are given in Figure~\ref{fig:wilson coeff constr}. For comparison, we display other recent constraints in Table \ref{tab:const comp}. As usual, the LHC constraints are still several orders of magnitude less stringent than the constraints from the electron EDM such that the observation of a deviation would imply a cancellation by a few orders of magnitude for the SMEFT contribution to the EDM. Our estimated constraints are similar in both processes for $\mathcal{O}_{\widetilde{W}WW}$ and lie between the semi-hadronic $WZ$ and VBF measurements by ATLAS. On the contrary, $W\gamma$ gives much better constraints on $\mathcal{O}_{\phi\widetilde{W}B}$ than $WZ$ and they are competitive with VBF. As a result, they could be use to confirm or disprove the deviation seen in this process. 

While our results do not seem much better than previous results, they can be improved in several ways. First, we have considered only one leptonic channel and adding the other leptonic channels will increase the statistics. For $WZ$, our observable is independent of the charge of the Z decay product and therefore suggest that it could be used also for its hadronic decay. Secondly, all the four processes can be combined. Thirdly, the NLO correction increase the SM cross-sections by roughly a factor 2. It this is true also for the SMEFT contributions to our observables, this could further enhance the sensitivity. Moreover, for the large contribution with an extra radiation, the jet could be used to better approximate the quark direction. Finally, the cross-sections are sufficiently large either to use differential asymmetries or to cut the phase space in order to improve the sensitivities.

\begin{table}[]
    \centering
    \begin{tabular}{c|c|c|c}
       Operators & $\sigma(pp\rightarrow l\nu jj)$ \cite{ATLAS:2017luz} & $\Delta \phi_{jj}$ \cite{Aad:2020sle} & EDM \cite{Panico:2018hal} \\
        \hline
        $\mathcal{O}_{\widetilde{W}WW}$ & [-14, 14] (expected) & [-0.12, 0.12] (expected) & $\leq 1.74~10^{-4}$ \\
                  & [-11, 11] (measured) &  [-0.11, 0.14] (measured) &  \\
        $\mathcal{O}_{\phi\widetilde{W}B}$ & // & [-1.06, 1.06] (expected) & $\leq 5.57~10^{-6}$ \\
                  & // &  [-0.23, 2.34] (measured) &  \\
    \end{tabular}
    \caption{Collection of the constraints on the two dimension-six operators with $\Lambda$ = 1TeV at 95\% CL. }
    \label{tab:const comp}
\end{table}

%% file: conclusion.tex
Our aim is to explore which observables could shed some light or constrain the presence of new sources of CP violation.
From the SMEFT Lagrangian in the Warsaw basis, we counted 6 CP-odd hermitian operators composed exclusively of bosonic fields and 1343 CP-odd non-hermitian operators with 3 generations. In order to reduce the number of operators and only keep the leading CP violating contributions, we neglected the masses of the light fermions and focus on the operators interference with the SM. As a result, the list of CP-odd operators which have leading contributions at high energy colliders is reduced to 10 operators under the $U(1)^{14}$ symmetry: the 6 bosonic operators and 4 operators containing the top quark. If the $U(1)^{13}$ symmetry is assumed instead, the total number of CP-odd operators increases to 17 with a massive bottom quark. 

Since the interference between the SM and the CP-odd operators is expected to have no contribution in CP-even observables, interference effects from CP-odd operators are phase space suppressed and require dedicated differential observables. We use the method developed in Ref.~\cite{Degrande:2020tno} to compare the efficiencies of various asymmetries in reviving those interferences in diboson production.

In particular, the asymmetries in the triple product of the quark, electron and neutral boson momenta with various approximation for the unmeasurable quark momentum were found to be the very efficient and to be more efficient than asymmetries based on previously proposed observables.  
In $W^\pm Z$, the efficiency of the best observable, $p_\perp(p_e, p_\sum)$, is 50\% for $\OWWW$ and 70\% for $\OBW$. In $W^\pm\gamma$, there is not one best observable in both channels even if $p_\perp(p_e, p_\sum)$ offers a middle ground option. For $W^+\gamma$, $p_\perp(p_e, p_\gamma)$ gives the largest asymmetries for $\OBW$ while $\Delta\varphi_{e\gamma}$ gives the largest asymmetries in all the other cases. For $W\gamma$, the efficiencies are all above 80\%.
Additionnally, the asymmetries based on $\sin{\phi_{WZ}}$ ~\cite{Azatov:2019xxn} displays good result for $\OWWW$ in $WZ$, especially in high-energy bins where they become more efficient than our triple product. However, this observable is almost blind to contributions from $\OBW$. Therefore, we could disentangle the two operators in this process by measuring both $\sin{\phi_{WZ}}$ and $p_\perp(p_e, p_\sum)$. The two variables are also complementary in term of EFT validity once the UV-theory is fixed as they get their sensitivities from different parts of the energy spectrum.

Since the direction of the initial quarks is very influential to build effective asymmetries, the leptonic colliders, even if their center-of-mass energy is lower, could have a good sensitivity as the direction of the initial leptons is known. 

As a conclusion, we have found simple observables that have a high efficiencies to revive the interferences between the CP-odd operators and the SM. They can be used as a starting point for more advanced CPV analyses but we also argue for their direct measurements as they are independent of any NP assumption.

\section*{Acknowlegments} We thanks J.-M. G\'erard for his insightful discussions and suggestions on this work. This work was funded by the F.R.S.-FNRS through the MISU convention F.6001.19. Computational resources have been provided by the supercomputing facilities of the Université Catholique de Louvain (CISM/UCL) and the Consortium des Équipements de Calcul Intensif en Fédération Wallonie Bruxelles (CÉCI) funded by the Fond de la Recherche Scientifique de Belgique (F.R.S.-FNRS) under convention 2.5020.11 and by the Walloon Region.